\documentclass[
aps,
pre,
amsmath,amssymb,amsfonts,
reprint,
floatfix,
nofootinbib,
]{revtex4-2}

\usepackage{graphicx}
\usepackage{dcolumn}
\usepackage{bm}

\usepackage{mathptmx}
\usepackage{xcolor}

\usepackage[caption=false]{subfig}
\usepackage{ragged2e}
\DeclareCaptionJustification{justified}{\justifying}

\usepackage{anyfontsize}

\usepackage{amsthm}

\newtheoremstyle{my_theorem_style_remark}{2.5pt}{2.5pt}{}{}{\slshape\bfseries}{.}{  }{\thmname{#1}\thmnumber{ #2}}
\theoremstyle{my_theorem_style_remark}
\newtheorem{remark}{Remark}
\newtheoremstyle{my_theorem_style}{2.5pt}{2.5pt}{\itshape}{}{\bfseries}{.}{  }{}
\theoremstyle{my_theorem_style}

\allowdisplaybreaks

\begin{document}
\title{Many interacting particles in solution. II. Screening-ranged expansion of electrostatic forces}

\author{Sergii V. Siryk}
\email{accandar@gmail.com}
\affiliation{CONCEPT Lab, Istituto Italiano di Tecnologia, Via E. Melen 83, 16152, Genova, Italy}
\author{Walter Rocchia}
\email[W. Rocchia \emph{(corresponding co-author)}: ]{walter.rocchia@iit.it}
\affiliation{CONCEPT Lab, Istituto Italiano di Tecnologia, Via E. Melen 83, 16152, Genova, Italy}

\begin{abstract}
\noindent
We present a fully analytical integration of the Maxwell stress tensor and derive exact relations for interparticle forces in systems of multiple dielectric spheres immersed in a polarizable ionic solvent, within the framework of the linearized Poisson--Boltzmann theory. Building upon the screening-ranged (in ascending orders of Debye screening) expansions of the potentials developed and rigorously analyzed in the accompanying works \textcolor{red}{\cite{supplem_pre,supplem_pre_math,supplem_prl}}, we construct exact screening-ranged many-body expansions for electrostatic forces in explicit analytical form. These results establish a rigorous foundation for evaluating screened electrostatic interactions in complex particle systems and provide direct analytical connections to, and systematic improvements upon, various earlier approximate or limited-case formulations available in the literature, both at zero and finite ionic strength.
\end{abstract}

\maketitle
\section{Introduction}
\label{section-Intro}
\noindent
The Poisson--Boltzmann equation (PBE)~\cite{Blossey2023,BesleyACR,JPCBSIYR,Cisneros_ChemRev} is one of the most reliable and widely used tools for estimating electrostatic quantities in biomolecular systems immersed in electrolytic solutions. It governs the mean-field electrostatic potential generated by fixed charges in a dielectric medium containing mobile ions, thus providing an effective description of the electrostatic interactions that steer physicochemical phenomena in systems with ionic atmospheres. Apart from biomolecular sciences, PBEs are actively utilized in colloidal sciences, plasma physics, and related fields (see \textcolor{red}{\cite{supplem_pre}} for a more detailed overview). The general PBE is a nonlinear partial differential equation~\cite{NakovMat,GS_2018,Cisneros_ChemRev}, and its solution poses considerable challenges even numerically. These difficulties persist, to a lesser extent, in its linearized form (LPBE). The latter, simpler equation—often referred to as the Debye--Hückel (DH) equation—is more manageable both numerically and analytically and provides a reasonable approximation for systems with low potentials~\cite{our_jcp,our_jpcb,Derb2,Derb4,BMP22,LM1,YDZ,KMMT2018}. In the case of highly charged solutes, the corresponding electric field sources can also be properly renormalized (see overviews in~\cite{Triz1,Boon2015,SchlaichHolm,supplem_pre,our_jcp,our_jpcb} and references therein). Consequently, the LPBE—or the even coarser electrostatic part of the DLVO (Derjaguin–Landau–Verwey–Overbeek) theory~\cite{KJSHS,NikLeeWas,FinBar2016,BesleyACR,C5NR04274G}—remains practically useful even in such cases. These considerations make the LPBE one of the most commonly employed tools for studying electrostatics in biomolecular and colloidal systems. Correspondingly, theoretical and computational research on LPBE-related topics continues to develop actively, generating a steady flow of work from both communities (see, e.g.~recent papers~\cite{our_jcp,our_jpcb,Yu3,Yu2021,BritoDenton,Tabrizi1,FilStar_jpcb,BesleyACR,Derb2,Derb4,DerbFil2017,LinQS,Jha1,AddisonSmithCooper2023,Silva1,SCW,BoWangjcp,RMDP,FinBar2016,MAR1,KMRR,BSBBC,FBYJBH,Ether2018,WK1,WK2,CoopJPCB,FolOnu,BMP22,Ponce2024,JNQS,GLCBB_2025,DiFlorio2025NextGenPB,FEM-BEM_diffuse_interface} and references therein).

While LPBE is most often used to estimate the electrostatic contribution to the total energy, the same framework also allows one to calculate electrostatic forces accurately, in principle (even though this can be challenging because the electric field may need to be evaluated on the molecular surface~\cite{AddisonSmithCooper2023,HBF_jctc_2017}). Several approaches have been developed for this purpose~\cite{AddisonSmithCooper2023,Gilson_et_al_1993,CheDzu2008,Derb2,Derb4,LuCheng,FilStar_jpcb,CC1993,CC1994,CYL2012,XCYWL2013,WMV22}, including the virtual displacement method (taking the gradient of the energy), the energy functional variation method, and the Maxwell stress tensor (MST) approach. A concise overview of these is provided in a recent paper~\cite{AddisonSmithCooper2023}. It is also shown in \cite{AddisonSmithCooper2023}, through numerical benchmarks within the LPBE framework, that the MST approach yields the most accurate results among various alternatives when tested on the same setups. The MST approach within the LPBE framework has been extensively studied over the last three decades (see, for example,~\cite{Derb2,Derb4,LuCheng,FilStar_jpcb,CC1993,CC1994,CYL2012,XCYWL2013,AddisonSmithCooper2023} and references therein). It is a versatile method that can be applied not only to conventional biomolecular systems with fixed charge distributions and transmission-type boundary conditions on solute-solvent interfaces but also to systems with other types of boundary conditions, such as fixed potentials~\cite{SC1999,Derb4,FilStar_jpcb} or linear charge regulation~\cite{CC1993,CC1994}. We also adopt the MST approach in the present study. 

Nevertheless, beyond abstract integral relations, explicit analytical expressions for the MST components on molecular surfaces—and the corresponding force expressions—have so far remained unavailable for general many-sphere dielectric systems within the LPBE framework. These expressions would be key for fast numerical simulations of polarizable sphere systems and for improving molecular dynamics simulations and models/codes~\cite{AddisonSmithCooper2023,DuanGan2025}. Only certain limited-case results have been reported previously (e.g.~for two-sphere systems with special symmetries, or systems at zero ionic strength; see below for a more detailed overview). This paper thus aims at bridging this gap, namely:
\newcounter{my_temp_list_counter0}
\begin{list}{\textbf{\arabic{my_temp_list_counter0}.} }{\usecounter{my_temp_list_counter0} \topsep=0pt \parsep=0pt \itemsep=0pt \partopsep=0pt \parskip=0pt \itemindent=0pt \leftmargin=0pt \labelwidth=0pt \labelsep=0pt}
    \item\label{nov_list_force_general}We derive fully analytical, exact, and general expressions for all components of electrostatic forces acting on arbitrary particles, modeled as interacting dielectric spheres embedded in an electrolytic environment. These results generalize and improve upon previously available analytical formulations~\cite{ClercxBossis1993,CC1993,CC1994,Derb2,Derb4,FengZhouWang,BBKS,FilStar_jpcb} (see Sec.~\ref{general_expressions_forces_sec} for details). In particular, we demonstrate that our analytical electrostatic theory and the resulting many-body force expressions reproduce, as a special case, the theoretical model proposed in~\cite{BBKS} describing exact interactions between a pair of monopolar dielectric spheres in vacuum at zero ionic strength (i.e.~governed by the Poisson rather than the Poisson–Boltzmann equation). This model has been widely discussed in subsequent studies~\cite{LCSB,GAMR2025}, particularly in the context of like-charge attraction and opposite-charge repulsion arising from polarization.
    \item\label{nov_list_energy_force_expansions}Furthermore—and we regard this as the most significant milestone of the present work—we combine the general analytical many-body force expressions derived in Point~\ref{nov_list_force_general} with the theory of screening-ranged expansions of potentials developed in the accompanying studies~\textcolor{red}{\cite{supplem_pre,supplem_prl}}. Based on the screening-ranged (organized by ascending Debye screening orders) expansion of solvent potential coefficients thoroughly constructed in the joint work~\textcolor{red}{\cite{supplem_pre}}, we build exact expansions—also ordered by increasing Debye screening—of all electrostatic force components acting on arbitrary particles, up to arbitrary screening order. The corresponding analytical construction is presented in Sec.~\ref{general_expansions_pot_force}. This provides, to our knowledge for the first time, an analytical formalism at the LPBE level that yields exact quantification of arbitrary-order Debye screening force components in many-body systems in explicit form, without imposing restrictive assumptions on system parameters (such as small particle radii, large interparticle separations, or small ratios thereof, which are common in prior literature). As an example, we show in Sec.~\ref{Appendix_screened_potential_coefficients_point_force_section} how the double-screened force expressions from our new many-body formalism generalize and improve upon the approximate analytical corrections to DLVO-type forces reported in~\cite{Derb2}.
\end{list}

The present article is the second part of a companion series: the first part \textcolor{red}{\cite{supplem_pre}} focuses on screening-ranged expansions for electrostatic potentials and energies, whereas here we address the corresponding forces. The theoretical foundation of the screening-ranged expansions of potential coefficients used in Point~\ref{nov_list_energy_force_expansions} relies on the spectral analysis of the associated composite Neumann--Poincar\'e-type operators, which is developed in detail in the companion paper~\textcolor{red}{\cite{supplem_pre_math}}.

\section{General construction of exact screening-ranged expansions of potentials and energies}
\label{statements_basics_section}
\subsection{Boundary value problem statement}
\label{basic_problem_statement0}
\noindent
Let us consider a general system consisting of $N$ non-overlapping spherical dielectric particles (represented by balls $\Omega_i$, $i=\overline{1,\ldots,N}$, where $\Omega_i\subset\mathbb R^3$ is mathematically considered as an open set in $\mathbb R^3$) immersed into the surrounding medium (electrolytic solvent -- e.g.~water and mobile ions) described by dielectric constant $\varepsilon_\text{sol}$ and Debye screening length $\kappa^{-1}>0$. Let us also note that since in this study $\kappa$ is mathematically treated rather as a parameter (of the LPBE \eqref{Lin_eqs_lpb} -- see below), we do not focus on its precise meaning (see \cite{LBD2023,Kjellander_JCP_2016}). Each particle $\Omega_i$ is centered at $\mathbf x_i\in\mathbb R^3$ and is characterized by its dielectric constant $\varepsilon_i$ and radius~$a_i$. The electrostatic potential $\Phi_{\text{in},i}(\mathbf r)$ in $\Omega_i$ (i.e.~as $r_i < a_i$ where $r_i$ is the radial coordinate of $\mathbf r \in \mathbb{R}^3$ measured from $\mathbf x_i$ so that $r_i = \left\|\mathbf r_i\right\|$, $\mathbf r_i = \mathbf r - \mathbf x_i$) satisfies the PE, while the corresponding potential $\Phi_{\text{out},i}$ in solvent, due to the presence of the $i$-th particle, obeys the LPBE~\cite{our_jcp,our_jpcb}:
\begin{subequations}
\label{Lin_eqs}
\begin{align}
&\Delta\Phi_{\text{in},i}(\mathbf r)= - \rho_i^\text{f}(\mathbf r)/(\varepsilon_i \varepsilon_0), & & \mathbf r\in\Omega_i,\label{Lin_eqs_poisson} \\
&\Delta\Phi_{\text{out},i}(\mathbf r) - \kappa^2 \Phi_{\text{out},i}(\mathbf r) = 0, & & \mathbf r\in\Omega_\text{sol}\label{Lin_eqs_lpb}
\end{align}
\end{subequations}
for all $i=\overline{1,\ldots,N}$, where $\rho_i^\text{f}(\mathbf r)$ is a (free or fixed) charge density distribution supported inside the $i$-th particle, and the solvent domain $\Omega_\text{sol} \mathrel{:=} \mathbb R^3\setminus\bigcup_{i=1}^N\overline{\Omega}_i$ (where $\overline{\Omega}_i$ is the $\mathbb R^3$-closure of $\Omega_i$). Due to the superposition principle adopted in the DH description \cite{Fish,Derb2} the total self-consistent electrostatic potential $\Phi(\mathbf r)$ of the whole system is~then
\begin{equation}
\label{Lin_eq_tot_pot}
\Phi(\mathbf r) = \left[
\begin{aligned}
&\Phi_{\text{in},i}(\mathbf r),\quad \mathbf r\in\Omega_i,\\
&\Phi_{\text{out}}(\mathbf r) \mathrel{:=} \sum\nolimits_{i=1}^N \Phi_{\text{out},i}(\mathbf r),\quad \mathbf r\in\Omega_\text{sol},
\end{aligned}
\right.
\end{equation} 
while at the boundaries of media potential $\Phi(\mathbf r)$ is subject to some boundary conditions (BCs) -- for example~\textcolor{red}{\cite{supplem_pre}}
\begin{subequations}
\label{Lin_eq_standard_bc}
\begin{gather}
\left.\Phi_{\text{in},i}\right|_{r_i \to a_i^-} = \left.\Phi_{\text{out}}\right|_{r_i \to a_i^+}, \label{Lin_eq_standard_bc_1st} \\
\varepsilon_i\left.(\mathbf n_i\cdot\nabla\Phi_{\text{in},i})\right|_{r_i\to a_i^-} - \varepsilon_\text{sol}\left.(\mathbf n_i\cdot\nabla\Phi_{\text{out}})\right|_{r_i\to a_i^+} = \sigma_i^\text{f}/\varepsilon_0 \label{Lin_eq_standard_bc_2nd}
\end{gather}
\end{subequations}
for all $i=\overline{1,\ldots,N}$, where $\mathbf n_i$ is the outer unit normal and $\sigma_i^\text{f}$ is a (free or fixed) charge density (if any) on the boundary $\partial\Omega_i$ ($r_i=a_i$) of $\Omega_i$, and $r_i\to a_i^{\pm}$ denotes approaching $\partial\Omega_i$ from exterior($+$)/interior($-$) of that particle. 

Potential $\Phi_{\text{in},i}$ admits \cite{our_jcp,our_jpcb} decomposition 
\begin{equation}
\label{Phi_in_i_decomposition}
\Phi_{\text{in},i} = \Hat\varPhi_{\text{in},i} + \Tilde\Phi_{\text{in},i}, 
\end{equation}
where $\Hat\varPhi_{\text{in},i}(\mathbf r) = \frac{1}{4\pi\varepsilon_0\varepsilon_i}\!\int_{\Omega_i}\!\frac{\rho_i^\text{f}(\mathbf r') \, d \mathbf r'}{\|\mathbf r - \mathbf r' \|}$ is the given particular solution to \eqref{Lin_eqs_poisson} representing the standard Coulombic potential in uniform infinite space generated by $\rho_i^\text{f}$, while $\Tilde\Phi_{\text{in},i}$ fulfills the Laplace equation, $\Delta\Tilde\Phi_{\text{in},i}=0$. We also point out the conventional conditions for potentials satisfying system \eqref{Lin_eqs}, ensuring physical feasibility~\cite{our_jcp,Jack}, namely $|\Tilde\Phi_{\text{in},i}|<\infty$ as $r_i\to0^+$ and $\Phi_{\text{out},i}\to0$ as $r_i\to+\infty$.

\subsection{Representation of potentials through their spherical Fourier coefficients}
\label{spherical_Fourier_coeffs_subsec}
\noindent
We seek for the potentials $\Tilde\Phi_{\text{in},i}(\mathbf r)$ and $\Phi_{\text{out},i}(\mathbf r)$ in the form~\textcolor{red}{\cite{supplem_pre_math}}
\begin{subequations}
\label{Lin_eqs_Phi_in_out}
\begin{gather}
\Tilde\Phi_{\text{in},i}(\mathbf r) = \sum\nolimits_{n,m}L_{n m,i} \Tilde r_i^n Y_n^m(\Hat{\mathbf r}_i),\label{Lin_eqs_Phi_in} \\
\Phi_{\text{out},i}(\mathbf r) = \sum\nolimits_{n,m}G_{n m,i}k_n(\Tilde r_i) Y_n^m(\Hat{\mathbf r}_i)\label{Lin_eqs_Phi_out}
\end{gather}
\end{subequations}
with some numeric coefficients $\{L_{n m,i}\}$ and $\{G_{n m,i}\}$; these coefficients are unknown and yet to be determined from BCs~\eqref{Lin_eq_standard_bc}. Here $i=\overline{1,\ldots,N}$, the scaled dimensionless radial variable $\Tilde r_i \mathrel{:=} \kappa r_i$, the sum $\sum_{n,m} \mathrel{:=} \sum_{0\le|m|\le n} =  \sum_{n=0}^{+\infty} \sum_{m=-n}^n$ (also notation $0\le\left|m\right|\le n$ for indices $n$ and $m$ henceforth means the running $n=\overline{0,\ldots,+\infty}$, $m=\overline{-n,\ldots,n}$), and unit vector $\Hat{\mathbf r}_i \mathrel{:=} \mathbf r_i/r_i$. Spherical angles $\theta_i$ and $\varphi_i$ emanating from $\Hat{\mathbf r}_i$ and then used in (complex-valued) spherical harmonics 
$Y_n^m$ (see definition \eqref{Ynm_definition}), are measured for any $i$ relative to a local coordinate system with center at $\mathbf x_i$ and axes parallel to those of some unique (fixed) global coordinate system. Also $k_n(x) \mathrel{:=} \sqrt{2/\pi}K_{n+1/2}(x)/\sqrt{x}$ and $i_n(x) \mathrel{:=} \sqrt{\pi/2}I_{n+1/2}(x)/\sqrt{x}$ are modified spherical Bessel functions of the 2nd and 1st kind (see Appendix~\ref{appendix_bessel_functions_summary}). Note that the spherical harmonics in relations \eqref{Lin_eqs_Phi_out} for different particles refer to the local coordinate frames of the corresponding particles (see the joint paper \textcolor{red}{\cite{supplem_pre}} for a more detailed explanation).

It can be derived (see detailed calculations in the joint work~\textcolor{red}{\cite{supplem_pre}}) that boundary conditions \eqref{Lin_eq_standard_bc} lead to the following matrix linear systems governing the coefficients of~\eqref{Lin_eqs_Phi_out}:
\begin{equation}
\label{eqs_G_intermediate_}
\mathsf A_i \Tilde{\mathbf G}_i + \sum\nolimits_{j=1,\, j\ne i}^N \mathsf B_{i j}\Tilde{\mathbf G}_j = \mathbf S_i ,\quad \forall i=\overline{1,\ldots,N},
\end{equation}
where we have introduced (infinite-size) column-vectors $\Tilde{\mathbf G}_i$, $\mathbf S_i$ and matrices $\mathsf A_i$, $\mathsf B_{i j}$, such that 
\begin{equation}
\label{various_matrix_definitions_0}
\begin{aligned}
\Tilde{\mathbf G}_i &\mathrel{:=} \left\{\Tilde G_{n m,i} \right\}_{n m} , \\ 
\mathbf S_i &\mathrel{:=} \left\{(2 n+1)\varepsilon_i\Hat L_{n m,i}/\Tilde a_i^{n+2} \right\}_{n m} ,\\
\mathsf A_i &\mathrel{:=} \operatorname{diagonal}\left\{\alpha_n(\Tilde a_i,\varepsilon_i) \Upsilon_{n,i}\right\}_{n m} , \\
\mathsf B_{i j} &\mathrel{:=} \left\{\beta_{n m, L M}(\Tilde a_i,\varepsilon_i,\mathbf R_{i j})\Upsilon_{L,j}\right\}_{n m, L M} 
\end{aligned}
\end{equation}
(index subscripts $n m$ and $L M$ mean pairs of indices $0\le | m |\le n$ and $0\le | M |\le L$ which enumerate rows and columns, respectively) with 
\begin{subequations}
\label{alpha_beta_definitions}
\begin{align}
& \alpha_n(\Tilde a_i,\varepsilon_i) \mathrel{:=} (\varepsilon_i-\varepsilon_\text{sol}) n \frac{k_n(\Tilde a_i)}{\Tilde a_i} + \varepsilon_\text{sol} k_{n+1}(\Tilde a_i), \label{alpha_definition} \\
&\begin{aligned}
\beta_{n m, L M}(\Tilde a_i,\varepsilon_i,\mathbf R_{i j}) \mathrel{:=} & \Bigl(\!\!(\varepsilon_i\!-\!\varepsilon_\text{sol}) n \frac{i_n(\Tilde a_i)}{\Tilde a_i} \!-\! \varepsilon_\text{sol} i_{n+1}(\Tilde a_i)\!\!\Bigr) \\ 
& \times \mathcal H_{n m}^{L M}(\mathbf R_{i j}) \, .
\end{aligned} \label{beta_definition}
\end{align}
\end{subequations}
Here, $\mathbf R_{i j} \mathrel{:=} \mathbf x_j-\mathbf x_i$ points from $\mathbf x_i$ to $\mathbf x_j$, $R_{i j} = \|\mathbf R_{i j}\|$, $\Hat{\mathbf R}_{i j} \mathrel{:=} \mathbf R_{i j}/R_{i j}$, $\Tilde R_{i j} \mathrel{:=} \kappa R_{i j}$, $\Tilde a_i \mathrel{:=} \kappa a_i$, and the specially-built quantities $\mathcal H_{n m}^{L M}(\mathbf R_{i j})$ take care of re-expansions of screened harmonics between local coordinate frames for correct imposing the boundary conditions~\textcolor{red}{\cite{supplem_pre}}: $\mathcal H_{l_1 m_1}^{L M}(\mathbf{R}_{i j}) \mathrel{:=} \sum_{l_2,m_2}\! (-1)^{l_1+l_2} H_{l_1 m_1 l_2 m_2}^{L M} k_{l_2}(\Tilde R_{i j}) Y_{l_2}^{m_2}(\Hat{\mathbf R}_{i j})$, where $H_{l_1 m_1 l_2 m_2}^{L M} \mathrel{:=} C_{l_1 0 l_2 0}^{L 0} C_{l_1 m_1 l_2 m_2}^{L M} \sqrt{\tfrac{4\pi (2 l_1+1) (2 l_2+1)}{2 L+1}}$ and $C_{l_1 m_1 l_2 m_2}^{L M} = \left<l_1 l_2; m_1 m_2 \mid L M\right>$ are Clebsch-Gordan coefficients. In \eqref{eqs_G_intermediate_} and \eqref{various_matrix_definitions_0}, for the mathematical convenience (see \textcolor{red}{\cite{supplem_pre}}) we have also introduced the scaling of the coefficients of \eqref{Lin_eqs_Phi_out}, namely 
\begin{equation}
\label{G_nm_scaling}
G_{n m,i} = \Tilde G_{n m,i} \Upsilon_{n,i} \quad\text{with}\quad \Upsilon_{n,i} \mathrel{:=} \left((2 n+1) k_n(\Tilde a_i) a_i\right)^{-1}.
\end{equation}

In \textcolor{red}{\cite{supplem_prl,supplem_pre}} we have constructed screening-ranged expansions 
\begin{equation}
\label{G_componentwise0_}
\Tilde{\mathbf G}_i = \sum_{\ell=0}^{+\infty}\Tilde{\mathbf G}_i^{(\ell)} 
\end{equation}
for the coefficients $\Tilde{\mathbf G}_i$ of potentials~\eqref{Lin_eqs_Phi_out}. Here superscript $\ell$ indicates the order of screening by Debye screening factors: $\Tilde{\mathbf G}_i^{(0)}$ is independent of any $R_{i j}$, $\Tilde{\mathbf G}_i^{(1)}\propto\frac{e^{-\kappa R_{i j}}}{R_{i j}}$ (with indices $j\ne i$), $\Tilde{\mathbf G}_i^{(2)}\propto\frac{e^{-\kappa R_{i k}}}{R_{i k}}\frac{e^{-\kappa R_{k j}}}{R_{k j}}$ (with $j$ and $k$ such that $i\ne k$, $k\ne j$), and so on -- see the joint paper \textcolor{red}{\cite{supplem_pre}} for the detailed construction of vectors $\Tilde{\mathbf G}_i^{(\ell)}$ and their element-wise representations $\Tilde{\mathbf G}_i^{(\ell)} = \{\Tilde G^{(\ell)}_{n m,i}\}_{n m}$ (for the readers' convenience, the necessary information is briefly recapitulated in Appendix~\ref{G_appendix_nm_ell_elementwise}). In turn, relation \eqref{G_componentwise0_} will pave the way for us to construct the relevant screening-ranged expansions also for the electrostatic forces (see Sec.~\ref{general_expansions_pot_force}).

\section{Exact quantification of interaction forces and their expansions in ascending orders of Debye screening factors}
\label{explicit_forces_section}
\subsection{General analytical force relations in multi-sphere systems}
\label{general_expressions_forces_sec}
\noindent
In general, the (total) electrostatic force $\mathbf F_i$ acting on the $i$-th particle due to the presence of all other particles can be expressed by integrating the normal component of Maxwell stress tensor \cite{Derb2,Derb4,LuCheng,FilStar_jpcb,CC1993,CC1994,CYL2012,XCYWL2013,WMV22,DTB2014}, namely
\begin{equation}
\label{Force_i_total}
\mathbf F_i = \oint_{\partial\Omega_i} \mathbf T_{\mathbf n_i} d s , \qquad \mathbf T_{\mathbf n_i} \mathrel{:=} \Breve{\mathbf T}_{\mathbf n_i} + \mathring{\mathbf T}_{\mathbf n_i} ,
\end{equation}
where the normal component of the (conventional) MST is $\Breve{\mathbf T}_{\mathbf n_i} = \bigl(E_{\mathbf n_i} \mathbf E - \frac{1}{2} (\mathbf E\cdot\mathbf E)\mathbf n_i \bigr)\varepsilon_0 \varepsilon_\text{sol} = \bigl(\frac{1}{2}(E_{\mathbf n_i}^2-\mathbf E_{\pmb{\tau}_i}\cdot\mathbf E_{\pmb{\tau}_i})\mathbf n_i + E_{\mathbf n_i} \mathbf E_{\pmb{\tau}_i}\bigr)\varepsilon_0 \varepsilon_\text{sol}$, electric field $\mathbf E$ is taken on the $i$-th particle's surface ($\mathbf E = \left.-\nabla\Phi_{\text{out}}\right|_{r_i\to a_i^+}$), $E_{\mathbf n_i}$ is the normal component of $\mathbf E$ (so that one has orthogonal decomposition $\mathbf E = E_{\mathbf n_i} \mathbf n_i + \mathbf E_{\pmb{\tau}_i}$). The osmotic pressure term besides the conventional MST, accounting for the difference in the local osmotic pressure from that in the bulk electrolyte~\cite{CC1994}, is $\mathring{\mathbf T}_{\mathbf n_i} = -\frac{1}{2}\varepsilon_0 \varepsilon_\text{sol}\kappa^2 \left.\Phi_{\text{out}}^2\right|_{r_i\to a_i^+} \mathbf n_i$. Using the MST-based approach for expressing the forces also appears to be beneficial in numerical calculations -- see a very recent paper \cite{AddisonSmithCooper2023} which benchmarks and compares different approaches for calculating electrostatic forces (such as the conventional energy gradient/virtual displacement method, energy functional variation approach, etc.), and concludes that the MST approach actually yields the most accurate results. A remarkable thing we will derive is that in the case of many spherical particles under the LPBE framework the MST normal component can be integrated fully analytically (see Appendix~\ref{appendix_force_derivation}), so that the $(x,y,z)$-components of force \eqref{Force_i_total} so obtained can then be actually expressed~as
\begin{equation}
\label{force_i}
(\mathbf{F}_i)_x = \operatorname{Re} \aleph_i , \qquad
(\mathbf{F}_i)_y = \operatorname{Im} \aleph_i , \qquad
(\mathbf{F}_i)_z = \beth_i,  
\end{equation}
where it is denoted 
\begin{subequations}
\label{aleph_beth_definitions}
\begin{align}
\aleph_i \mathrel{:=} &\, \varepsilon_0\varepsilon_\text{sol}\sum_{n,m}\sqrt{\tfrac{(n-m+1) (n-m+2)}{(2 n+1) (2 n+3)}} \bigl( B_{n+1,m-1;i} A_{n,m;i}^\star \label{aleph_ordinary}\\ 
& - \Tilde a_i^2 \Psi_{n+1,m-1;i}\Psi_{n,m;i}^\star\bigr) , \label{aleph_osmotic}\\ 
\beth_i \mathrel{:=} &\, \varepsilon_0\varepsilon_\text{sol}\sum_{n,m} \sqrt{\tfrac{(n-m+1) (n+m+1)}{(2 n+1) (2 n+3)}} \bigl(B_{n+1,m;i} A_{n,m;i}^\star \label{beth_ordinary}\\
& - \Tilde a_i^2\Psi_{n+1,m;i}\Psi_{n,m;i}^\star\bigr) ,\label{beth_osmotic}
\end{align}
\end{subequations}
where
\begin{align*}
A_{n,m;i} & \mathrel{:=} \Xi_{n,m;i}-n\Psi_{n,m;i}, \quad B_{n,m;i} \mathrel{:=} \Xi_{n,m;i}+(n+1)\Psi_{n,m;i} ,\\
\Psi_{n,m;i} & \mathrel{:=} k_n(\Tilde a_i) G_{n m,i} + i_n(\Tilde a_i)\sum_{j=1,\, j\ne i}^N \; \sum_{L,M}\mathcal H_{n m}^{L M}(\mathbf R_{i j}) G_{L M,j}, \\
\Xi_{n,m;i} & \mathrel{:=} \bigl(n k_n(\Tilde a_i) - \Tilde a_i k_{n+1}(\Tilde a_i)\bigr) G_{n m,i} \\ 
& + \bigl(n i_n(\Tilde a_i) + \Tilde a_i i_{n+1}(\Tilde a_i)\bigr)\sum_{j=1,\, j\ne i}^N \; \sum_{L,M}\mathcal H_{n m}^{L M}(\mathbf R_{i j}) G_{L M,j} . 
\end{align*}
The derivation of relations \eqref{force_i}-\eqref{aleph_beth_definitions} is provided in Appendix~\ref{appendix_force_derivation}. Note that force components \eqref{force_i} are entirely expressed through the coefficients $\{G_{n m,i}\}$ of DH potentials \eqref{Lin_eqs_Phi_out} and are obviously real -- indeed, $\beth_i$ in \eqref{force_i} can also be rewritten as $\beth_i = \operatorname{Re}\Bigl( \varepsilon_0\varepsilon_\text{sol}  \sum\limits_{n=0}^{+\infty} \, \sum\limits_{m=0}^n \sqrt{\frac{(n-m+1) (n+m+1)}{(2 n+1) (2 n+3)}}(2-\delta_{m,0})\bigl( B_{n+1,m;i} A_{n,m;i}^\star - \Tilde a_i^2\Psi_{n+1,m;i}\Psi_{n,m;i}^\star \bigr)\!\Bigr)$. 

By virtue of \eqref{Force_i_total}, total force $\mathbf F_i$ naturally admits general representation 
\begin{equation}
\label{Force_i_decomposition_general}
\mathbf F_i = \Breve{\mathbf F}_i + \mathring{\mathbf F}_i ,
\end{equation}
where $\mathring{\mathbf F}_i$ is caused by $\mathring{\mathbf T}_{\mathbf n_i}$ which apparently vanishes at zero ionic strength ($\kappa=0$) and whose contribution is expressed by $\Psi\Psi^\star$-terms in the equalities for $\aleph_i$ and $\beth_i$ (i.e.~$\Psi_{n+1,m-1;i}\Psi_{n,m;i}^\star$ in $\aleph_i$ and $\Psi_{n+1,m;i}\Psi_{n,m;i}^\star$ in $\beth_i$ -- see \eqref{aleph_osmotic}, \eqref{beth_osmotic}). In turn, addend $\Breve{\mathbf F}_i$ stems from $\Breve{\mathbf T}_{\mathbf n_i}$ (whose contribution is expressed by~\eqref{aleph_ordinary},~\eqref{beth_ordinary}). Let us note that particular explicit forms of $(\Breve{\mathbf{F}}_i)_z$ in the case of a system consisting of \emph{two} $z$-aligned dielectric spheres and assuming the total azimuthal symmetry (AS) in the problem were recently obtained in papers \cite{Derb2,Derb4,FengZhouWang} (note that for such an AS configuration one has $(\mathbf{F}_i)_x=(\mathbf{F}_i)_y=0$, and only ($m=0$)-terms survive in $(\mathbf{F}_i)_z$); the osmotic pressure contributions $(\mathring{\mathbf{F}}_i)_z$ in such system were additionally quantified in a very recent work~\cite{FilStar_jpcb}. 

Force relations \eqref{force_i}-\eqref{aleph_beth_definitions} established above and the screening-ranged expansion \eqref{G_componentwise0_} for potential coefficients in turn give rise to build the screening-ranged expansions of electrostatic forces (see Sec.~\ref{general_expansions_pot_force} below).
\begin{remark}
\label{remark_force_representation_via_legendre}
In many studies (see e.g.~\cite{our_jcp,our_jpcb,Derb2,Derb4,FilStar_jpcb}) potentials $\Phi_{\text{out},i}$ (see \eqref{Lin_eqs_Phi_out}) use real harmonics ($P_n^m(\cos\theta_i)\cos(m\varphi_i)$, $P_n^m(\cos\theta_i)\sin(m\varphi_i)$) rather than complex-valued spherical harmonics $Y_n^m(\Hat{\mathbf r}_i)$ for representing angular dependence, that~is $\Phi_{\text{out},i}(\mathbf r) = \sum\limits_{n=0}^{+\infty}\frac{K_{n+1/2}(\Tilde r_i)}{\sqrt{\Tilde r_i}}\sum\limits_{m=0}^n\bigl( G_{n m,i}' \cos(m\varphi_i) + H_{n m,i}' \sin(m\varphi_i)\bigr)P_n^m(\cos\theta_i)$ with some real numeric coefficients $G_{n m,i}'$ and $H_{n m,i}'$; with the solvent potentials so represented force relations \eqref{force_i} recast to the following form (as one can see, much less concise as compared to expressions \eqref{force_i}-\eqref{aleph_beth_definitions} which relied on the representation of potentials through complex-valued harmonics):
\begin{subequations}
\label{force_Leg_}
\begin{flalign}
& (\mathbf{F}_i)_x = \pi\varepsilon_0\varepsilon_\text{sol}\sum_{\chi\in\{G',\,H'\}} \; \sum_{m=m_\chi}^{+\infty}(1+\delta_{m,0}) \sum_{n=m+1}^{+\infty} \frac{1}{2 n+1} && \notag \\ 
&\ \times \Biggl[\biggl(\frac{B_{n+1,m;i}^\chi A_{n,m+1;i}^{\chi}}{2 n+3} - \frac{B_{n,m+1;i}^{\chi} A_{n-1,m;i}^{\chi}}{2 n-1}\biggr) && \label{force_x_Leg_ord}\\
&\ - \Tilde a_i^2 \Psi_{n,m+1;i}^{\chi}\biggl(\frac{\Psi_{n+1,m;i}^\chi}{2 n+3} - \frac{\Psi_{n-1,m;i}^\chi}{2 n-1}\biggr)
 \Biggr] \frac{(n+m+1)!}{(n-m-1)!}\, ,\!\!\!\!\! && \label{force_x_Leg_osm} \\
& (\mathbf{F}_i)_y = \pi\varepsilon_0\varepsilon_\text{sol}\!\sum_{\chi\in\{G',\,H'\}}\!(-1)^\chi \!\sum_{m=m_\chi}^{+\infty}\!(1+\delta_{m,0})\! \sum_{n=m+1}^{+\infty}\! \frac{1}{2 n+1} && \notag \\ 
&\ \times \Biggl[\biggl(\frac{B_{n+1,m;i}^\chi A_{n,m+1;i}^{\Bar\chi}}{2 n+3} - \frac{B_{n,m+1;i}^{\Bar\chi} A_{n-1,m;i}^{\chi}}{2 n-1}\biggr) && \label{force_y_Leg_ord}\\
&\ - \Tilde a_i^2 \Psi_{n,m+1;i}^{\Bar\chi}\biggl(\frac{\Psi_{n+1,m;i}^\chi}{2 n+3} - \frac{\Psi_{n-1,m;i}^\chi}{2 n-1}\biggr)
 \Biggr] \frac{(n+m+1)!}{(n-m-1)!}\, ,\!\!\!\!\! && \label{force_y_Leg_osm} \\
& (\mathbf{F}_i)_z = 2\pi\varepsilon_0\varepsilon_\text{sol}\sum_{\chi\in\{G',\,H'\}} \; \sum_{m=m_\chi}^{+\infty} \sum_{n=m+1}^{+\infty} \Biggl[B_{n,m;i}^\chi A_{n-1,m;i}^\chi \!\! && \label{force_z_Leg_ord}\\
&\ - \Tilde a_i^2 \Psi_{n-1,m;i}^{\chi}\Psi_{n,m;i}^{\chi}
 \Biggr] \frac{(1+\delta_{m,0}) (n+m)!}{(4 n^2-1) (n-m-1)!}\, , && \label{force_z_Leg_osm}
\end{flalign}
\end{subequations}
where denoted $m_\chi \mathrel{:=} \left\{\!\begin{smallmatrix} 0, & \text{if } \chi=G' \\ 1, & \text{if } \chi=H' \end{smallmatrix}\!\right\}$, $(-1)^\chi \mathrel{:=} \left\{\!\begin{smallmatrix} +1, & \text{if } \chi=G' \\ -1, & \text{if } \chi=H' \end{smallmatrix}\!\right\}$, $\Bar\chi \mathrel{:=} \left\{\!\begin{smallmatrix} H', & \text{if } \chi=G' \\ G', & \text{if } \chi=H' \end{smallmatrix}\!\right\}$, 
$A_{n,m;i}^\chi \mathrel{:=} \Xi_{n,m;i}^\chi-n\Psi_{n,m;i}^\chi$, $B_{n,m;i}^\chi \mathrel{:=} \Xi_{n,m;i}^\chi+(n+1)\Psi_{n,m;i}^\chi$. Quantities  $\Xi_{n,m;i}^\chi$ and $\Psi_{n,m;i}^\chi$ represent $\Phi_{\text{out}}$ and the normal component of electric field on $\partial\Omega_i$ (expressed this time through real harmonics, see~\eqref{def_Psi_Ynm}): $\left.\Phi_{\text{out}}\right|_{r_i\to a_i^+} = \sum\limits_{n=0}^{+\infty}\sum\limits_{m=0}^n\bigl(\Psi_{n,m;i}^{G'} \cos(m\varphi_i) + \Psi_{n,m;i}^{H'} \sin(m\varphi_i)\bigr)P_n^m(\cos\theta_i)$ and $E_{\mathbf n_i} = \left. -\frac{\partial\Phi_{\text{out}}}{\partial r_i}\right|_{r_i\to a_i^+} = \frac{-1}{a_i}\sum\limits_{n=0}^{+\infty}\sum\limits_{m=0}^n\bigl(\Xi_{n,m;i}^{G'} \cos(m\varphi_i) + \Xi_{n,m;i}^{H'} \sin(m\varphi_i)\bigr)P_n^m(\cos\theta_i)$; using \eqref{def_Psi_Ynm} and \eqref{eqs_Ynm_Pnm_} these coefficients $\Psi_{n,m;i}^{G'}$, $\Psi_{n,m;i}^{H'}$, $\Xi_{n,m;i}^{G'}$, $\Xi_{n,m;i}^{H'}$ can straightforwardly be expressed via the defined above coefficients $\Psi_{n,m;i}$ and $\Xi_{n,m;i}$ (see \eqref{aleph_beth_definitions}) that underpinned the force expressions when the counterpart potential expansions through complex-valued harmonics were employed.

Expressions \eqref{force_x_Leg_ord}, \eqref{force_y_Leg_ord} and \eqref{force_z_Leg_ord} represent components of $\Breve{\mathbf{F}}_i$, whereas expressions \eqref{force_x_Leg_osm}, \eqref{force_y_Leg_osm} and \eqref{force_z_Leg_osm} represent components of $\mathring{\mathbf{F}}_i$. In the particular case of a two-sphere system ($N=2$) with spheres' centers lying on the axis $\mathbf z$ and the assumption of AS, relations \eqref{force_z_Leg_ord} and \eqref{force_z_Leg_osm} can be immediately recast to recover the expressions obtained in recent papers \cite{Derb2,Derb4} and \cite{FilStar_jpcb}, respectively; finally, the AS expressions derived in earlier works \cite{CC1993,CC1994} can also be eventually extracted from \eqref{force_z_Leg_ord}-\eqref{force_z_Leg_osm}. In the case of zero ionic strength ($\kappa=0$), analytical expressions for the force components in terms of multipolar moments were also obtained earlier in~\cite{ClercxBossis1993}.

\end{remark}
\begin{remark}
\label{remark_force_kappa_zero}
The obtained force relations \eqref{force_i}-\eqref{aleph_beth_definitions} correctly reproduce the forces also in the $\kappa\to0$ limit (so that the LPBE \eqref{Lin_eqs_lpb} turns into the simpler Poisson equation). Although the case of zero ionic strength goes beyond the main scope of this study and should be discussed in detail elsewhere, let us only observe here that \eqref{force_i}-\eqref{aleph_beth_definitions} are able to reproduce, as a very particular case, the theoretical model proposed in 2010 in the salient work \cite{BBKS} where the exact description of the interaction between a pair of monopolar dielectric spheres in vacuum was given. Interestingly, this model was extensively referenced and discussed in subsequent works (see e.g.~\cite{LCSB} and recent \cite{GAMR2025}), especially within the context of capturing like-charge attraction / opposite-charge repulsion effects due to polarization. Indeed, in \cite{BBKS} it was derived that in the system of two $z$-aligned dielectric spheres ($\mathbf x_1 = (0,0,0)$, $\mathbf x_2 = (0,0,R)$, $R>0$) bearing free charges $q_1$ and $q_2$ uniformly distributed over the respective spherical surfaces (i.e.~$\sigma_i^\text{f} = q_i/(4\pi a_i^2) = \text{const}$) the electrostatic force is exactly expressed, adhering to our notations,~by 
\begin{align}
& F_{1 2} = 4\pi\varepsilon_0\varepsilon_\text{sol}\sum_{l=0}^{+\infty}\mathfrak G_{l,1}\mathfrak G_{l+1,1}\frac{(l+1)(k_1+1)+1}{(k_1-1)a_1^{2 l+3}} \label{force_kappa_zero_BBKS_0}\\
& = \frac{-q_1 q_2}{4\pi\varepsilon_0\varepsilon_\text{sol} R^2} + q_1\sum_{m=1}^{+\infty} \, \sum_{l=0}^{+\infty}\mathfrak G_{l,1}\frac{(k_2-1)m(m+1)}{(k_2+1) m +1}\frac{(m+l)!}{m! l!} \notag\\
& \times \frac{a_2^{2 m+1}}{R^{2 m+l+3}} + 4\pi\varepsilon_0\varepsilon_\text{sol}\sum_{l=1}^{+\infty}\frac{(l+1)(k_1+1)+1}{(k_1-1)a_1^{2 l+3}}\mathfrak G_{l,1}\mathfrak G_{l+1,1},\notag
\end{align}
where $k_i=\varepsilon_i/\varepsilon_\text{sol}$, $\varepsilon_\text{sol}=1$ (vacuum), and the multipole moment coefficients $\{\mathfrak G_{l,1}\}$ yield the pertinent azimuthally symmetric expansion $\Phi_{\text{out},1}(\mathbf r) = \sum_{l=0}^{+\infty}\mathfrak G_{l,1} r_1^{-l-1} P_l(\cos\theta_1)$. We adopted in \eqref{force_kappa_zero_BBKS_0} the sign convention in which $F_{1 2}$ is negative for a repulsion force acting on sphere~1 (the original relations derived in \cite{BBKS}, see \cite[Eqs.~(25)-(31)]{BBKS}, come with the reverted signs). Then it can be shown that in this particular two-sphere situation, general many-body relations \eqref{force_i}-\eqref{aleph_beth_definitions} can be recast to yield relation \eqref{force_kappa_zero_BBKS_0} in the limit $\kappa\to0$ (this passage, however, requires some rather fine calculations which are postponed to Appendix~\ref{force_kappa_zero_BBKS_Appendix_derivation}). While the starting addend $\frac{-q_1 q_2}{4\pi\varepsilon_0\varepsilon_\text{sol} R^2}$ in \eqref{force_kappa_zero_BBKS_0} represents the regular Coulombic force (as if the particles were non-polarizable or represented point charges $q_1$ and $q_2$ separated by a distance $R$), the terms beyond it can capture like-charge attraction / opposite-charge repulsion effects -- this can be especially easily illustrated in the simple situation where, for instance, sphere~1 is non-polarizable or represents a point charge (see \cite{BBKS}), so that $\mathfrak G_{l,1} = \frac{q_1 \delta_{l,0}}{4\pi\varepsilon_0\varepsilon_\text{sol}}$ ($\delta_{\cdot,\cdot}$ is a Kronecker delta), therefore $$F_{1 2} = \frac{-q_1 q_2}{4\pi\varepsilon_0\varepsilon_\text{sol} R^2} + \frac{q_1^2 (k_2-1)}{4\pi\varepsilon_0\varepsilon_\text{sol} R^2}\sum\limits_{m=1}^{+\infty}\frac{m(m+1) a_2^{2 m+1}}{((k_2+1)m+1) R^{2 m+1}}$$ from which it is now clear that the second term becomes attractive/repulsive depending only on the sign of $k_2-1$ (and, furthermore, it is also easy to see how to regulate the system parameters so that the absolute value of this term exceeds that of the first, Coulombic, term).
\end{remark}
\subsection{General construction of expansions (in ascending orders of Debye screening factors) for forces}
\label{general_expansions_pot_force}
\noindent
The main inconvenience in using the general force relations \eqref{force_i}-\eqref{aleph_beth_definitions} (as well as their various simplified/limited-case predecessors obtained in earlier works and mentioned above) is that the DH potentials' coefficients $\{G_{n m,i}\}$ are unknown and must be found from \eqref{eqs_G_intermediate_} \emph{before} actually using \eqref{force_i}-\eqref{aleph_beth_definitions} --- indeed, having solved the set of equations \eqref{eqs_G_intermediate_} the coefficients $\{G_{n m,i}\}$ must then be substituted into relations \eqref{force_i}-\eqref{aleph_beth_definitions} to derive the force. This prevents the possibility of exact explicit a-priori calculation of forces / derivation explicit analytical expressions for them and has therefore hitherto been considered the main obstacle to the effective use of force relations of this type (see e.g.~the very recent paper \cite{GAMR2025} for additional comments on relevant issues). Hence, we remove this obstacle by means of equality \eqref{G_componentwise0_}, which expresses the solution to system \eqref{eqs_G_intermediate_} as a rapidly converging screening-ranged expansion whose addends are resolved explicitly in terms of the basic system parameters. Namely, combining the above-derived force relations \eqref{force_i}-\eqref{aleph_beth_definitions} with relation \eqref{G_componentwise0_} gives rise to build the desired screening-ranged force expansion:
\begin{gather}
\mathbf{F}_i = \sum_{\ell=1}^{+\infty}\mathbf{F}_i^{(\ell)}, \label{Force_expansion_vec}\\ 
(\mathbf{F}_i^{(\ell)})_x = \operatorname{Re} \aleph_i^{(\ell)} , \quad 
(\mathbf{F}_i^{(\ell)})_y = \operatorname{Im} \aleph_i^{(\ell)} , \quad
(\mathbf{F}_i^{(\ell)})_z = \beth_i^{(\ell)} \notag
\end{gather}
with quantifying all its addends exactly and explicitly --- indeed, the corresponding $\ell$-screened components are expressed~by
\begin{align*}
&\aleph_i^{(\ell)} = \varepsilon_0\varepsilon_\text{sol}\!\sum\nolimits_{n,m}\!\sqrt{\tfrac{(n-m+1) (n-m+2)}{(2 n+1) (2 n+3)}} \sum\nolimits_{k=0}^\ell \bigl( B_{n+1,m-1;i}^{(k)} A_{n,m;i}^{(\ell-k)}{}^\star \\
&\quad - \Tilde a_i^2 \Psi_{n+1,m-1;i}^{(k)}\Psi_{n,m;i}^{(\ell-k)}{}^\star\bigr) \quad\text{and} \\ 
&\beth_i^{(\ell)} = \varepsilon_0\varepsilon_\text{sol}\sum\nolimits_{n,m} \sqrt{\tfrac{(n-m+1) (n+m+1)}{(2 n+1) (2 n+3)}} \sum\nolimits_{k=0}^\ell \bigl(B_{n+1,m;i}^{(k)} A_{n,m;i}^{(\ell-k)}{}^\star \\
&\quad - \Tilde a_i^2\Psi_{n+1,m;i}^{(k)}\Psi_{n,m;i}^{(\ell-k)}{}^\star\bigr), \intertext{where $\forall k=\overline{0,\ldots,\ell}$ it is defined}
&A_{n,m;i}^{(k)} = \Xi_{n,m;i}^{(k)} - n\Psi_{n,m;i}^{(k)}, \qquad B_{n,m;i}^{(k)} = \Xi_{n,m;i}^{(k)} + (n+1)\Psi_{n,m;i}^{(k)} ,\\
&\Psi_{n,m;i}^{(k)} = k_n(\Tilde a_i) G_{n m,i}^{(k)} + i_n(\Tilde a_i)\!\sum\nolimits_{j=1, j\ne i}^N \! \sum\nolimits_{L,M}\mathcal H_{n m}^{L M}\!(\mathbf R_{i j}) G_{L M,j}^{(k-1)}\!, \\
&\Xi_{n,m;i}^{(k)} = \bigl(n k_n(\Tilde a_i) - \Tilde a_i k_{n+1}(\Tilde a_i)\bigr) G_{n m,i}^{(k)} + \bigl(n i_n(\Tilde a_i) + \Tilde a_i i_{n+1}(\Tilde a_i)\bigr)\\
&\quad\times \sum\nolimits_{j=1,\, j\ne i}^N\sum\nolimits_{L,M}\mathcal H_{n m}^{L M}(\mathbf R_{i j}) G_{L M,j}^{(k-1)}
\end{align*} 
(it is understood that sums involving $G_{L M,j}^{(k-1)}$ in $\Psi_{n,m;i}^{(k)}$ and $\Xi_{n,m;i}^{(k)}$ are omitted at $k=0$, and $G_{n m,i}^{(k)} = \Tilde G_{n m,i}^{(k)} \Upsilon_{n,i}$ owing to the scale $\Upsilon_{n,i}$ defined above, see~\eqref{G_nm_scaling}). Let us note that force $\mathring{\mathbf{F}}_i$ (see decomposition \eqref{Force_i_decomposition_general}) is expressed by terms of the $\Psi_\cdot^{(\cdot)}\Psi_\cdot^{(\cdot)}{}^\star$-type products in the above expressions $\aleph_i^{(\ell)}$ and $\beth_i^{(\ell)}$. The sum in \eqref{Force_expansion_vec} actually starts with $\ell=1$ -- indeed, detailed analysis of the expressions for $\aleph_i^{(\ell)}$ and $\beth_i^{(\ell)}$ affirms that quantities $\aleph_i^{(0)}$ and $\beth_i^{(0)}$ always vanish, see Appendix~\ref{non-screened-force-components-vanish-appendix}, hence $\mathbf{F}_i^{(0)}$ turns out to be zero. Let us also emphasize once again that the screening-ranged coefficients $\{G^{(\ell)}_{n m}\}$ here are supplied by the expansion~\eqref{G_componentwise0_} constructed in the joint works~\textcolor{red}{\cite{supplem_prl, supplem_pre}}. Similarly to what was observed in \textcolor{red}{\cite{supplem_pre}} in the case of electrostatic energy, numerical calculations reaffirm (see Sec.~\ref{section_num_modeling} below) the rapid convergence of the screening-ranged force expansions~\eqref{Force_expansion_vec} constructed above.
\begin{remark}
\label{remark_torque}
Although not considered in detail here, the same formalism allows for \emph{torque} derivation in cases of asymmetric charge distributions.
\end{remark}
\begin{widetext}
\subsection{Example: $N$ spheres with centrally-located point charges}
\label{Appendix_screened_potential_coefficients_point_force_section}
\noindent
For this special case, in Appendix~\ref{Appendix_screened_potential_coefficients_point_force} we provide expanded customized expressions for all auxiliary quantities involved in \eqref{Force_expansion_vec} up to and including the 2nd order of screening, from which the following force relations then easily follow.
\subsubsection*{Single-screened force ($\ell=1$)} 
\noindent
Denoting $\pmb{\mathfrak F}_i \mathrel{:=} \sum\limits_{j=1,j\ne i}^N \frac{q_j e^{\Tilde a_j}}{1+\Tilde a_j}\frac{e^{-\Tilde R_{i j}}}{\Tilde R_{i j}}\Bigl(1+\frac{1}{\Tilde R_{i j}}\Bigr)\left\{
\begin{smallmatrix}
\sin\theta_{i j}\cos\varphi_{i j} \\
\sin\theta_{i j}\sin\varphi_{i j} \\
\cos\theta_{i j}
\end{smallmatrix}
\right\} = \sum\limits_{j=1,j\ne i}^N \frac{q_j e^{\Tilde a_j}}{1+\Tilde a_j}\frac{e^{-\Tilde R_{i j}}}{\Tilde R_{i j}}\Bigl(1+\frac{1}{\Tilde R_{i j}}\Bigr) \Hat{\mathbf R}_{i j}$ (curly brackets apparently indicate $\left\{
\begin{smallmatrix}
x \\
y \\
z
\end{smallmatrix}\right\}$-components, $\theta_{i j}$ and $\varphi_{i j}$ are spherical angles of $\Hat{\mathbf R}_{i j}$), $d_{1,i}^\prime \mathrel{:=} \left((\varepsilon_i+2\varepsilon_\text{sol})(1+\Tilde a_i)+\varepsilon_\text{sol}\Tilde a_i^2\right)^{-1}$, $d_{1,i} \mathrel{:=} (\varepsilon_i+2\varepsilon_\text{sol}) d_{1,i}^\prime$, for the singly-screened ($\ell=1$) force we then obtain
\begin{equation}
\label{forces_single_screening_all_}
\begin{aligned}
& \Breve{\mathbf F}_i^{(1)}
= \frac{-q_i e^{\Tilde a_i} \kappa^2 d_{1,i}}{4\pi\varepsilon_0\varepsilon_\text{sol}} \pmb{\mathfrak F}_i , \quad
\mathring{\mathbf F}_i^{(1)}
= \frac{-q_i e^{\Tilde a_i} \Tilde a_i^2 \kappa^2 d_{1,i}^\prime}{4\pi\varepsilon_0 (1+\Tilde a_i)} \pmb{\mathfrak F}_i \quad \Longrightarrow \quad \mathbf F_i^{(1)} = \Breve{\mathbf F}_i^{(1)} + \mathring{\mathbf F}_i^{(1)}
= \frac{-q_i e^{\Tilde a_i} \kappa^2}{4\pi\varepsilon_0\varepsilon_\text{sol}(1+\Tilde a_i)} \pmb{\mathfrak F}_i ,
\end{aligned}
\end{equation}
so that $\mathbf F_i^{(1)}$ expectedly yields the familiar DLVO force expression, $\mathbf F_i^\text{DLVO}$. 
\subsubsection*{Double-screened force ($\ell=2$)}
\noindent
For the double-screened ($\ell=2$) force contributions we obtain the following expanded relations, previously unknown in the literature (excepting some approximate two-sphere particular results discussed below):
\begin{subequations}
\label{forces_double_screening_all_}
\begin{align}
&\begin{aligned}
& \!\!\! \Breve{\mathbf F}_i^{(2)} = \frac{d_{3,i} d_{4,i} \kappa^2}{4\pi\varepsilon_0\varepsilon_\text{sol}}\!\!\sum_{j=1, j\ne i}^N \; \sum_{p=1, p\ne i}^N \frac{q_j q_p e^{\Tilde a_j + \Tilde a_p} e^{-\Tilde R_{i j}} e^{-\Tilde R_{i p}}}{(1+\Tilde a_j) (1+\Tilde a_p) \Tilde R_{i j} \Tilde R_{i p}}\!\Bigl(\! 1 \!+\! \frac{1}{\Tilde R_{i p}} \Bigr) \Bigl(1+\frac{3}{\Tilde R_{i j}}+\frac{3}{\Tilde R_{i j}^2}\Bigr) \bigl(3(\Hat{\mathbf R}_{i j}\cdot\Hat{\mathbf R}_{i p})\Hat{\mathbf R}_{i j}-\Hat{\mathbf R}_{i p}\bigr) -\frac{q_i e^{\Tilde a_i} d_{1,i} \kappa^2}{8\pi\varepsilon_0\varepsilon_\text{sol}}\\
&\times\!\!\sum_{j=1, j\ne i}^N \; \sum_{p=1, p\ne j}^N \!\frac{q_p e^{\Tilde a_p} e^{-\Tilde R_{i j}}e^{-\Tilde R_{j p}}}{(1+\Tilde a_p)\Tilde R_{i j}\Tilde R_{j p}}\!\biggl[\!\Bigl(\! 1 \!+\!\frac{1}{\Tilde R_{i j}}\!\Bigr) d_{0,j} \Hat{\mathbf R}_{i j} \!+\! \Bigl\{\!\Hat{\mathbf R}_{j p}\!+\!\Bigl(\! 1 \!+\!\frac{3}{\Tilde R_{i j}}\!+\!\frac{3}{\Tilde R_{i j}^2}\!\Bigr)\!\bigl(3(\Hat{\mathbf R}_{i j}\!\cdot\!\Hat{\mathbf R}_{j p})\Hat{\mathbf R}_{i j}\!-\!\Hat{\mathbf R}_{j p}\!\bigr)\!\Bigr\}\!\Bigl(\! 1\!+\!\frac{1}{\Tilde R_{j p}}\!\Bigr) d_{2,j} \biggr]\!+\! \cdots\!,
\end{aligned} \label{monopoles_double_standard} \\
&\begin{aligned}
& \!\!\! \mathring{\mathbf F}_i^{(2)} = \frac{-\kappa^2\Tilde a_i^2}{4\pi\varepsilon_0} \!\!\sum_{j=1, j\ne i}^N \; \sum_{p=1, p\ne i}^N \frac{q_j q_p e^{\Tilde a_j + \Tilde a_p} e^{-\Tilde R_{i j}} e^{-\Tilde R_{i p}}}{(1+\Tilde a_j) (1+\Tilde a_p) \Tilde R_{i j} \Tilde R_{i p}}\! \biggl[ \frac{e^{2\Tilde a_i}\Tilde a_i d_{1,i}^\prime}{1+\Tilde a_i}\Bigl(1+\frac{1}{\Tilde R_{i j}}\Bigr) \Hat{\mathbf R}_{i j} + \varepsilon_\text{sol}d_{3,i}^\prime d_{4,i}^\prime\!\Bigl(\!1\!+\!\frac{1}{\Tilde R_{i p}}\!\Bigr) \Bigl(1+\frac{3}{\Tilde R_{i j}}+\frac{3}{\Tilde R_{i j}^2}\Bigr) \\
&\times \bigl(3(\Hat{\mathbf R}_{i j}\cdot\Hat{\mathbf R}_{i p})\Hat{\mathbf R}_{i j}-\Hat{\mathbf R}_{i p}\bigr) \biggr] - \frac{q_i e^{\Tilde a_i} d_{1,i}^\prime \kappa^2 \Tilde a_i^2}{8\pi\varepsilon_0(1+\Tilde a_i)}\!\!\sum_{j=1, j\ne i}^N \; \sum_{p=1, p\ne j}^N \frac{q_p e^{\Tilde a_p}}{1+\Tilde a_p} \frac{e^{-\Tilde R_{i j}}}{\Tilde R_{i j}}\!\frac{e^{-\Tilde R_{j p}}}{\Tilde R_{j p}}\!\!\biggl[\!\Bigl(\!1\!+\!\frac{1}{\Tilde R_{i j}}\!\Bigr) d_{0,j} \Hat{\mathbf R}_{i j} + \Bigl\{\! \Hat{\mathbf R}_{j p} \!+\! \Bigl(\! 1 \!+\!\frac{3}{\Tilde R_{i j}}\!+\!\frac{3}{\Tilde R_{i j}^2}\!\Bigr)\\
&\times\bigl(3(\Hat{\mathbf R}_{i j}\!\cdot\!\Hat{\mathbf R}_{j p})\Hat{\mathbf R}_{i j}\!-\!\Hat{\mathbf R}_{j p}\!\bigr)\! \Bigr\}\!\Bigl(\!1\!+\!\frac{1}{\Tilde R_{j p}}\!\Bigr) d_{2,j} \!\biggr] + \cdots .
\end{aligned} \label{monopoles_double_osmotic}
\end{align}
\end{subequations}
The expressions displayed in \eqref{forces_double_screening_all_} were calculated in an exact way and are originating simply from $n,L\le1$ addends of \eqref{Force_expansion_vec} (all the quantities participating in \eqref{Force_expansion_vec} are listed in Appendix~\ref{Appendix_screened_potential_coefficients_point_force} in their customized form tailored to this system); arbitrary higher-order terms beyond this range can also be easily deduced from explicit relations reported in full in Appendix~\ref{Appendix_screened_potential_coefficients_point_force}. Here we denoted shorthands $d_{0,i} \mathrel{:=} e^{2\Tilde a_i} \frac{\Tilde a_i-1}{\Tilde a_i+1}+1$, $d_{2,i} \mathrel{:=} e^{2\Tilde a_i}\frac{(\varepsilon_i+2\varepsilon_{\mathsf{sol}})(\Tilde a_i-1)-\varepsilon_{\mathsf{sol}}\Tilde a_i^2}{(\varepsilon_i+2\varepsilon_{\mathsf{sol}})(1+\Tilde a_i)+\varepsilon_{\mathsf{sol}}\Tilde a_i^2}+1 $, 
$d_{3,i}^\prime \mathrel{:=} \frac{e^{2\Tilde a_i}\Tilde a_i^3}{(\varepsilon_i+2\varepsilon_\text{sol})(1+\Tilde a_i)+\varepsilon_\text{sol}\Tilde a_i^2}$, $d_{4,i}^\prime \mathrel{:=} \frac{1}{(2\varepsilon_i+3\varepsilon_\text{sol})(\Tilde a_i^2+3\Tilde a_i+3)+\varepsilon_\text{sol}\Tilde a_i^2(1+\Tilde a_i)}$, $d_{3,i} \mathrel{:=} (\varepsilon_i-\varepsilon_\text{sol})d_{3,i}^\prime$, $d_{4,i} \mathrel{:=} (2\varepsilon_i+3\varepsilon_\text{sol}) d_{4,i}^\prime$.

\bigskip

\paragraph*{Instantiation to the two-sphere case.}For example, in the particular case of two ($N=2$) $z$-aligned spheres ($\mathbf x_1 = (0,0,0)$, $\mathbf x_2 = (0,0,R)$, $R>0$), $(\mathbf{F}_i)_x=(\mathbf{F}_i)_y=0$ and many-body relations~\eqref{forces_double_screening_all_} boil down to the following:
\begin{align}
&\!\!\!
\begin{aligned}
& (\Breve{\mathbf{F}}_i^{(2)})_z = \frac{q_i^2\kappa^2 e^{2\Tilde a_i - 2\Tilde R} d_{1,i}}{8\pi\varepsilon_0\varepsilon_\text{sol}(1+\Tilde a_i)\Tilde R^2}\!\Bigl(\!1\!+\!\frac{1}{\Tilde R}\!\Bigr)\!\biggl(\!\!\Bigl(3\!+\!\frac{6}{\Tilde R}\!+\!\frac{6}{\Tilde R^2}\!\Bigr)d_{2,j} - \!d_{0,j}\!\!\biggr)  + \frac{q_j^2\kappa^2 e^{2\Tilde a_j-2\Tilde R} d_{3,i} d_{4,i} }{2\pi\varepsilon_0\varepsilon_\text{sol}(1\!+\!\Tilde a_j)^2\Tilde R^2}\!\Bigl(\!1\!+\!\frac{1}{\Tilde R}\Bigr)\!\Bigl(\!1\!+\!\frac{3}{\Tilde R}\!+\!\frac{3}{\Tilde R^2}\Bigr) + \cdots ,
\end{aligned} \label{force_two_spheres_2scr_nonosmotic} 
\\
&\!\!\!
\begin{aligned}
& (\mathring{\mathbf{F}}_i^{(2)})_z = \frac{q_i^2\kappa^2 \Tilde a_i^2 e^{2\Tilde a_i - 2\Tilde R} d_{1,i}^\prime}{8\pi\varepsilon_0(1+\Tilde a_i)^2\Tilde R^2}\!\Bigl(\!1\!+\!\frac{1}{\Tilde R}\!\Bigr)\!\!\biggl(\!\!\Bigl(\!3\!+\!\frac{6}{\Tilde R}\!+\!\frac{6}{\Tilde R^2}\!\!\Bigr)d_{2,j} - \!d_{0,j}\!\!\biggr)\!  - \!\frac{q_j^2\kappa^2 e^{2\Tilde a_j-2\Tilde R} d_{3,i}^\prime}{4\pi\varepsilon_0(1\!+\!\Tilde a_j)^2\Tilde R^2}\!\Bigl(\!1\!+\!\frac{1}{\Tilde R}\!\Bigr)\!\biggl(\!\frac{1}{1+\Tilde a_i} \! + \! 2\varepsilon_\text{sol}\Tilde a_i^2 d_{4,i}^\prime \Bigl(\!1\!+\!\frac{3}{\Tilde R}\!+\!\frac{3}{\Tilde R^2}\!\Bigr)\!\!\biggr)\! + \cdots
\end{aligned} \notag 
\end{align}
as $i=1$, $j=2$; these relations in their current (non-simplified -- see below) form do not seem to be previously known to the best of our knowledge, although their further simplification recovers the formulas previously derived in recent work \cite{Derb2} -- namely, if in the above $(\Breve{\mathbf{F}}_i^{(2)})_z$ one assumes that $\varepsilon_i=\varepsilon_\text{sol}$ (sphere $i$ is non-polarizable), $q_j=0$ (sphere $j$ is polarizable with zero charge), $\Tilde a_i \ll \Tilde R \ll 1$ and $\Tilde a_j \ll \Tilde R \ll 1$ (weak screening and large inter-particle separation of small particles), then treating $\Tilde a_j$ as small parameter and hereby approximating $d_{2,j}$ and $d_{0,j}$ via 
$d_{2,j} \approx \frac{2(\varepsilon_j-\varepsilon_\text{sol})}{3(\varepsilon_j+2\varepsilon_\text{sol})}\Tilde a_j^3$, $d_{0,j} \approx \frac{2}{3}\Tilde a_j^3$ (i.e.~by the leading terms of Taylor expansions of $d_{2,j}$ and $d_{0,j}$ in $\Tilde a_j$~\textcolor{red}{\cite{supplem_pre}}), one then immediately obtains from~\eqref{force_two_spheres_2scr_nonosmotic} 
\begin{equation}
\label{Derb_jcp_eq34}
(\Breve{\mathbf{F}}_i^{(2)})_z \approx \frac{q_i^2 a_j^3 e^{-2\kappa R}(1+\kappa R)}{4\pi\varepsilon_0\varepsilon_\text{sol} R^5}\biggl(\frac{\varepsilon_j-\varepsilon_\text{sol}}{\varepsilon_j+2\varepsilon_\text{sol}}(2+2\kappa R+\kappa^2 R^2)-\frac{1}{3}\kappa^2 R^2\biggr) ,
\end{equation}
which exactly recovers the result of \cite[Eq.~(34)]{Derb2} derived there under the above particular restrictions when approximating ion-molecule interaction (with sphere $i$ representing a non-polarizable ion and sphere $j$ representing a neutral molecule), and in this case the physical interpretation of different terms of simplified expression \eqref{Derb_jcp_eq34} can roughly be given as follows (see \cite[Sec.~III~A~2]{Derb2} for details): the 2nd term represents the electrolyte polarization around the molecule due to its finite size (note that this term does not vanish as $\varepsilon_j=\varepsilon_\text{sol}$ but nullifies in the absence of screening as $\kappa\to0$), while the 1st one can be derived from the interaction energy between a point charge and a dipole in ionized medium with the polarizability of molecule $j$ defined through the Clausius-Mossotti relation.
\end{widetext}
\section{Numerical modeling}
\label{section_num_modeling}
\begin{figure*}
\includegraphics[trim=0.1cm 0.15cm 0.1cm 0.1cm,clip=true,width=0.98\linewidth]{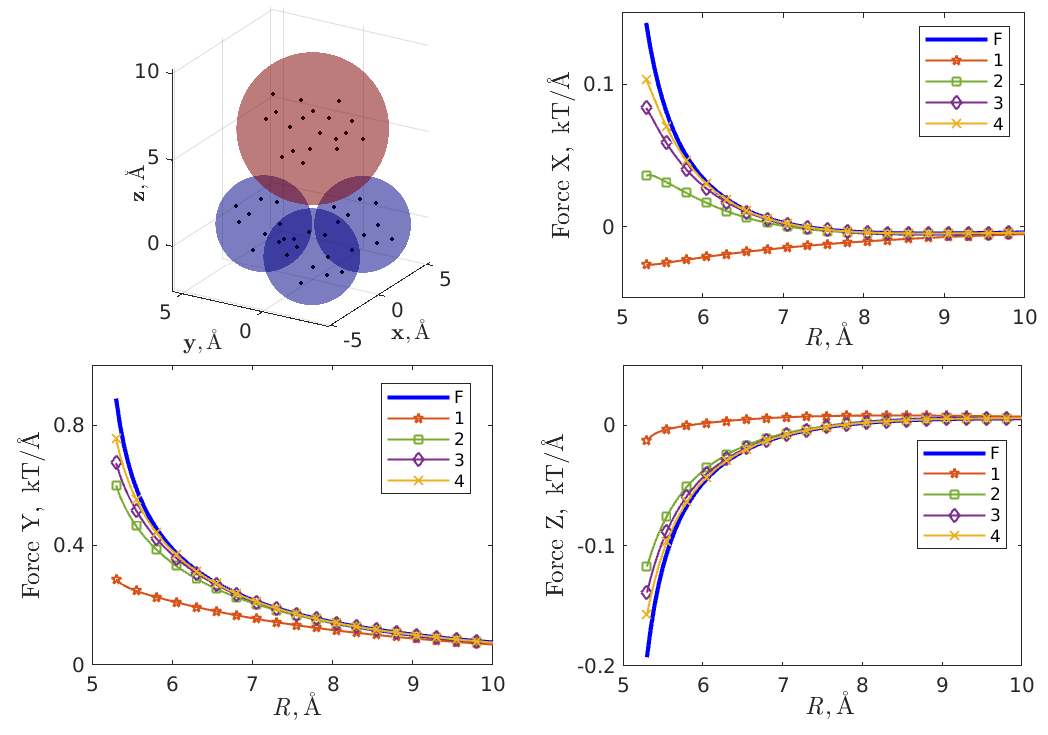}
\caption{The subfigure in the upper left corner shows the arrangement of spheres at $R=5.3$~\AA. The remaining subfigures show the Cartesian components of the total electrostatic force exerted on sphere~3 depending on $R$ which varies from $5.3$~\AA~to $10$~\AA~with step $0.05$~\AA~(continuous lines are used to guide the eye): Line F depicts the full (true) $\mathbf F_i$ for $i=3$ (see \eqref{force_i}), while Lines 1, 2, 3, and 4 illustrate the convergence of screening-ranged force expansion \eqref{Force_expansion_vec} depicting $\mathbf F_i^{(1)}$, $\mathbf F_i^{(1)}+\mathbf F_i^{(2)}$, $\mathbf F_i^{(1)}+\mathbf F_i^{(2)}+\mathbf F_i^{(3)}$, and $\mathbf F_i^{(1)}+\mathbf F_i^{(2)}+\mathbf F_i^{(3)}+\mathbf F_i^{(4)}$, respectively. (Here, $\mathrm k$ is the Boltzmann constant, $\mathrm T$ is the temperature.)}
\label{ArgGlu_forces_sphere3}
\end{figure*}
\begin{figure}
\includegraphics[trim=0.25cm 0.0cm 0.95cm 0.45cm,clip=true,width=0.99\linewidth]{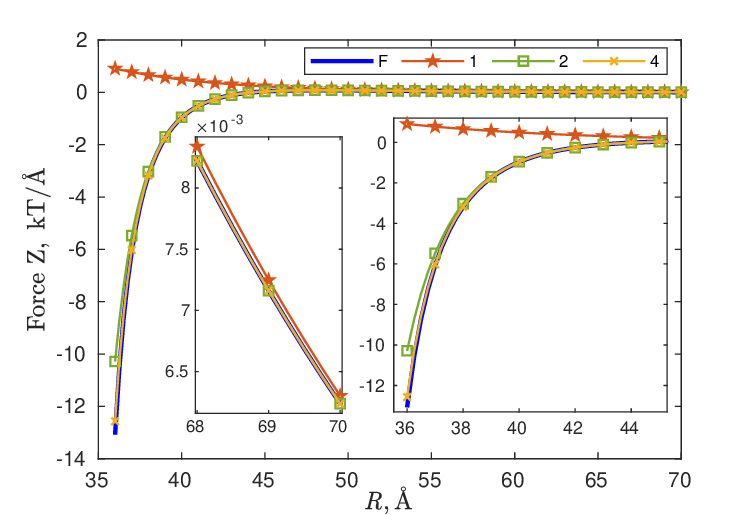}
\caption{Force depending on intercenter distance $R$, acting on sphere~2 in the two-sphere system. (Line legend is as in Fig.~\ref{ArgGlu_forces_sphere3}; embedded insets show close-up views.)}
\label{temp_attr_3_2_25_force}
\end{figure}
\noindent

To illustrate the convergence of the screening-ranged force expansion constructed in~\eqref{Force_expansion_vec}, we first consider two representative examples previously used in our joint study~\textcolor{red}{\cite{supplem_pre}} to benchmark the convergence of screening-ranged energy expansions. All parameter settings for the numerical modeling, including the cutoff parameter $n_\text{max}=25$, are taken from that work.

In the first example, we examine a system of four interacting spheres containing arginine and glutamate charge distributions. Specifically, the spheres (see Fig.~\ref{ArgGlu_forces_sphere3}) are centered at $\mathbf x_1=\bigl(-\frac{R}{2},-\frac{R}{2\sqrt{3}},0\bigr)$, $\mathbf x_2=\bigl(\frac{R}{2},-\frac{R}{2\sqrt{3}},0\bigr)$, $\mathbf x_3=\bigl(0,\frac{R}{\sqrt{3}},0\bigr)$, and $\mathbf x_4=\bigl(0,0,R\frac{\sqrt{407}}{\sqrt{300}}\bigr)$, with radii $a_1=a_2=a_3=2.6$~\AA\ and $a_4=4.1$~\AA, and dielectric constants $\varepsilon_1=2$, $\varepsilon_2=3$, $\varepsilon_3=4$, $\varepsilon_4=5$. The solvent parameters—typical for biophysical systems—are $\varepsilon_\text{sol}=80$ (aqueous solution) and $\kappa^{-1}=8.071$~\AA, corresponding to $0.145$~M physiological NaCl concentration at room temperature ($25^{\circ}$C). The distances between $\mathbf x_1$, $\mathbf x_2$, and $\mathbf x_3$ are equal to $R$, while each of these centers is separated from $\mathbf x_4$ by $1.3R$. Each sphere contains a tightly inscribed fixed charge distribution, with the distance between the outermost charges and the sphere surface being less than $0.1$~\AA. These distributions correspond to clouds of point charges representing an arginine–glutamate pair modeled in~\cite{MS2013,our_jpcb} (CHARMM22 force field). Glutamate charge distributions are embedded in spheres~1--3, while the arginine charge distribution is embedded in sphere~4. Figure~\ref{ArgGlu_forces_sphere3} presents the dependence of the force acting on sphere~3 on the intercenter distance $R$, which varies from $5.3$~\AA\ (where the spheres are tightly packed so that the separation between surfaces is less than $0.2$~\AA, and the minimal surface separation among spheres~1--3 is $0.1$~\AA) up to $10$~\AA. As in the case of electrostatic energy analyzed in~\textcolor{red}{\cite{supplem_pre}}, the leading single-screened term $\mathbf F_i^{(1)}$ alone does not robustly approximate the total force $\mathbf F_i$, particularly at short separations. Adding the leading double-screened correction $\mathbf F_i^{(2)}$ significantly improves accuracy, while inclusion of higher-order screened correction addends $\mathbf F_i^{(3)}$ and $\mathbf F_i^{(4)}$ yields further refinement.

In the second example, we study a system of two spheres with $\varepsilon_1=\varepsilon_2=80$, $a_1=\frac{350}{11}$~\AA, and $a_2=\frac{35}{11}$~\AA, such that $a_1/a_2=10$ and $a_1+a_2=35$~\AA. Point charges $q_1=3e$ and $q_2=2e$ are located at the centers $\mathbf x_1=(0,0,0)$ and $\mathbf x_2=(0,0,R)$, respectively. The solvent dielectric constant is $\varepsilon_\text{sol}=2$, and $\kappa$ remains the same as in the previous example. In this azimuthally symmetric configuration, only the $z$-components of the forces are nonzero. The force acting on sphere~2 ($i=2$) as a function of the intercenter distance $R=R_{12}$, varying from 36~\AA\ to 70~\AA\ in 0.25~\AA\ increments, is shown in Fig.~\ref{temp_attr_3_2_25_force}. Like-charge attraction is observed at short separations ($R\lesssim45$~\AA, which is also consistent with the energy profile data reported in~\textcolor{red}{\cite{supplem_pre}}) -- this attraction is not captured by the leading DLVO term $\mathbf F_i^{(1)}$ but is accurately reproduced when the first correction addend $\mathbf F_i^{(2)}$ is included.

\begin{figure*}
\includegraphics[trim=1.46cm 0.75cm 1.87cm 0.7cm,clip=true,width=0.95\linewidth]{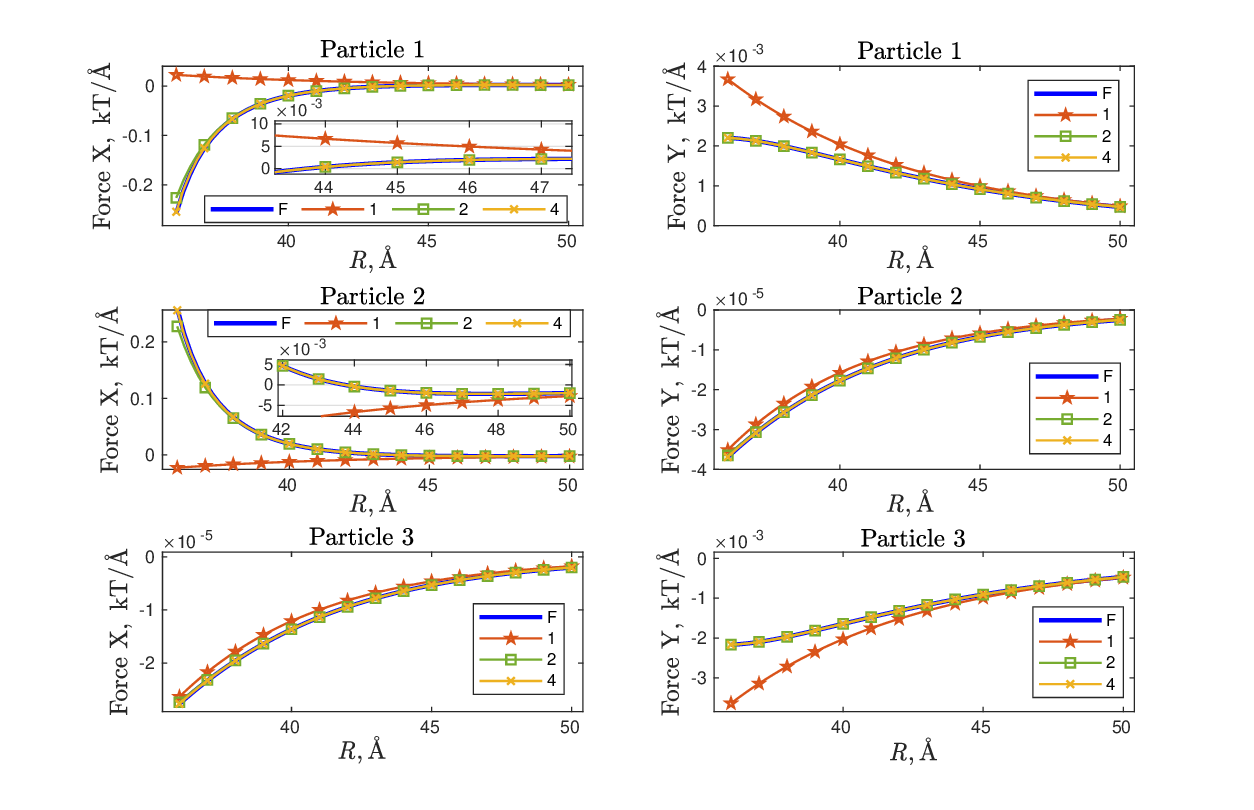}
\caption{Force components depending on distance $R$ varying from $36\text{~\AA}$ to $50\text{~\AA}$. (Line legends are as in Fig.~\ref{ArgGlu_forces_sphere3}; embedded insets show close-up views.)}
\label{temp_3_spheres_force}
\end{figure*}
\begin{figure}
\includegraphics[trim=1.97cm 0.2cm 2.68cm 0.65cm,clip=true,width=0.999\linewidth]{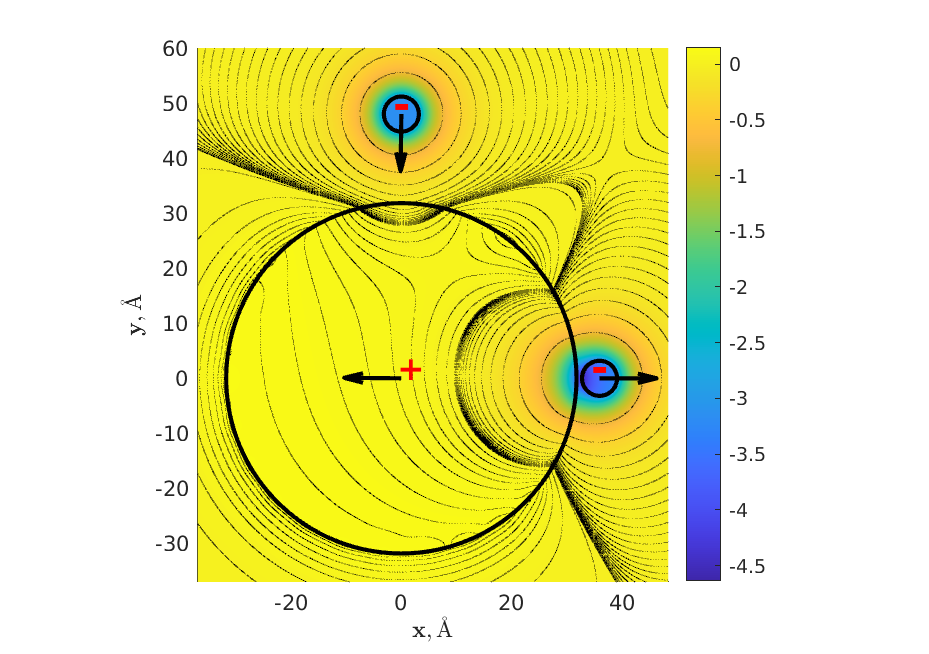}
\caption{Dimensionless (by dividing by $\mathrm k \mathrm T/e$) potential on the plane $z=0$ and its isolines at $R=36$~\AA. Arrows indicate the normalized (unit) directions of the forces acting on the particles.}
\label{temp_3_spheres_potential}
\end{figure}
Finally, in our third example we modify the previous one, namely we consider a 3-sphere system consisting of one larger sphere ($a_1=\frac{350}{11}$~\AA, $\varepsilon_1=2$) and two smaller spheres ($a_2=a_3=\frac{35}{11}$~\AA, $\varepsilon_2=\varepsilon_3=3$); $\varepsilon_\text{sol}=80$ and $\kappa$ remains the same as in the previous examples. The spheres carry fixed charges $q_1=3 e$ and $q_2=q_3=-2 e$ uniformly distributed over their surfaces (unlike a similar example considered in \textcolor{red}{\cite{supplem_prl}}, where a configuration with fixed centrally-located point charges was used), and are centered at $\mathbf x_1=(0,0,0)$, $\mathbf x_2=(R,0,0)$, and $\mathbf x_3=(0,R+12\text{\AA},0)$. Then Figs.~\ref{temp_3_spheres_force} and \ref{temp_3_spheres_potential} show the force components (note that in this planar arrangement only $x$- and $y$-components of forces are nonzero) and the potential, respectively; a short-range ($R\lesssim44$~\AA) opposite-charge repulsion between spheres 1 and 2 is observed and is accurately captured by~$\mathbf F_i^{(2)}$. 

Let us finally note that comparison of the convergence behavior of the screening-ranged force expansions in Figs.~\ref{ArgGlu_forces_sphere3} and~\ref{temp_attr_3_2_25_force} with the energy expansions reported in~\textcolor{red}{\cite{supplem_pre}} reveals that the force series generally converges more slowly than its energy counterpart. This slower convergence likely arises because differentiating a series of the form $\sum_{n\ge0} c_n \bigl(\frac{e^{-\Tilde R}}{R}\bigr)^n$ with respect to $R$ (here for the sake of simplicity of illustration we assume $N=2$ and $z$-aligned particles with characteristic separation $R$) introduces additional $n$-dependent factors --- specifically, applying $-\partial_R$ yields $(\kappa+R^{-1})\sum_{n\ge0} c_n n \bigl(\frac{e^{-\Tilde R}}{R}\bigr)^n$, thereby transforming the original ``energy'' series with coefficients $\{c_n\}$ into a ``force'' series with coefficients $\{c_n n\}$, which naturally converges more slowly.

\section{Summary and outlook}
\label{summary_conclusions}
\noindent
In this study, we investigated interparticle electrostatic forces in systems of interacting dielectric spherical particles immersed in a surrounding medium, modeled within the LPBE framework. We performed a fully analytical integration of the Maxwell stress tensor and derived general exact expressions~\eqref{force_i}-\eqref{aleph_beth_definitions} for the components of the force acting on an arbitrary particle in such many-body systems. These expressions generalize and improve upon the results of a number of earlier studies (see Sec.~\ref{general_expressions_forces_sec} for details). The purely electrostatic limit of zero ionic strength (i.e.~$\kappa \to 0$) is also correctly recovered; in particular, for azimuthally symmetric two-sphere systems, our exact analytical results reproduce those obtained in~\cite{BBKS}, which have been widely used in subsequent literature to explain the counterintuitive like-charge attraction and opposite-charge repulsion effects (see Remark~\ref{remark_force_kappa_zero}). Building upon these general expressions and the screening-ranged expansion of solvent potential coefficients developed in our companion work~\textcolor{red}{\cite{supplem_pre}} (see~\eqref{G_componentwise0_}), we further derived exact screening-ranged expansions for all components of the electrostatic force acting on a particle. The analytical formulation of this construction and explicit expressions for the screened force components are detailed in Sec.~\ref{general_expansions_pot_force}. To the best of our knowledge, this work presents the first explicit and fully many-body screening-ranged expansion of electrostatic forces available in the literature. Some simplified forms of the leading higher-order screened terms—specifically, the double-screened component $\Breve{\mathbf{F}}_i^{(2)}$—were previously derived in~\cite{Derb2} for the special case of two interacting particles ($N=2$) under additional restrictive assumptions (see~\eqref{Derb_jcp_eq34}). Closely related expressions for two-sphere systems under constant potential boundary conditions were also derived later in~\cite{Derb4}.

Finally, numerical modelling results presented in Sec.~\ref{section_num_modeling} confirm the rapid convergence of the screening-ranged force expansion~\eqref{Force_expansion_vec} developed in this study, further validating the accuracy and utility of the proposed analytical framework.

\section{Acknowledgments}
\label{sec:Acknowledgements}

We acknowledge the financial support from the European 
Union - NextGenerationEU and the Ministry of University and Research (MUR), 
National Recovery and Resilience Plan (NRRP): 
Research program CN00000013 “National Centre for HPC, 
Big Data and Quantum Computing”, CUP: J33C22001180001, funded by the D.D. n.1031, 17.06.2022 and Mission 4, Component 2, Investment 1.4 - Avviso “Centri Nazionali” - D.D. n. 3138, 16 December 2021.

\section*{Data availability}
\noindent
Codes/data can be found in~\textcolor{red}{\cite{our_github_rep}}.

\appendix

\section{Derivation of formulas~\eqref{force_i}-\eqref{aleph_beth_definitions}}
\label{appendix_force_derivation}
\noindent
The derivation of force equalities \eqref{force_i}-\eqref{aleph_beth_definitions} is not too complicated conceptually, but is rather tediously cumbersome; let us briefly describe the key points of this derivation. Operating in spherical coordinates associated with the $i$-th particle we can express $\Phi_{\text{out}}$, $E_{\mathbf n_i}$, and $\mathbf E_{\pmb{\tau}_i} =  E_{\Hat{\pmb\theta}_i}\Hat{\pmb\theta}_i + E_{\Hat{\pmb\varphi}_i}\Hat{\pmb\varphi}_i$ (here $\Hat{\pmb\theta}_i$ and $\Hat{\pmb\varphi}_i$ are the local angular unit spherical basis vectors~\cite{Jack}) on $\partial\Omega_i$ (see~\eqref{Force_i_total})~as 
\begin{align}
& \left.\Phi_{\text{out}}\right|_{r_i\to a_i^+} = \sum\limits_{n,m}\Psi_{n,m;i}Y_n^m(\Hat{\mathbf r}_i), \label{def_Psi_Ynm}\\
& E_{\mathbf n_i} = \left. -\frac{\partial\Phi_{\text{out}}}{\partial r_i}\right|_{r_i\to a_i^+} =\left|\text{use \eqref{diff_modifiedBessel}}\right| = \frac{-1}{a_i}\sum\limits_{n,m}\Xi_{n,m;i}Y_n^m(\Hat{\mathbf r}_i), \notag\\
& E_{\Hat{\pmb\theta}_i} = \!\left.\Bigl(\frac{-1}{r_i}\frac{\partial\Phi_{\text{out}}}{\partial\theta_i} \Bigr)\right|_{r_i\to a_i^+} = \frac{(1-\mu_i^2)^{1/2}}{a_i}\sum_{n,m} \Psi_{n,m;i}\frac{\partial Y_n^m(\Hat{\mathbf r}_i)}{\partial\mu_i} , \notag\\
& E_{\Hat{\pmb\varphi}_i} = \!\left.\Bigl(\frac{-1}{r_i \sin\theta_i}\frac{\partial\Phi_{\text{out}}}{\partial\varphi_i} \Bigr)\right|_{r_i\to a_i^+} = -\frac{1}{(1-\mu_i^2)^{1/2} a_i} \notag\\ 
& \quad \times \sum_{n,m} \Psi_{n,m;i}\frac{\partial Y_n^m(\Hat{\mathbf r}_i)}{\partial\varphi_i} , \notag
\end{align}
where $\mu_i \mathrel{:=} \cos\theta_i$ (note that $Y_n^m(\Hat{\mathbf r}_i) = Y_n^m(\theta_i,\varphi_i)$ in fact depends on $\mu_i$ rather than $\theta_i$ -- see definition \eqref{Ynm_definition}). Since potentials \eqref{Lin_eqs_Phi_out} are real~\textcolor{red}{\cite{supplem_pre_math}}, hereby $G_{n, -m,i} = (-1)^m G_{n m,i}^\star$ for all $i$, it is then easy to verify that also
\begin{align}
& \Xi_{n,-m;i} = (-1)^m \Xi_{n,m;i}^\star, \qquad \Psi_{n,-m;i} = (-1)^m \Psi_{n,m;i}^\star . \label{Xi_Psi_for_negative_m}
\intertext{Indeed, using the symmetry relation $\mathcal H_{n, -m}^{L, -M}(\mathbf{R}_{i j}) = (-1)^{m+M}\mathcal H_{n m}^{L M}(\mathbf{R}_{i j})^\star$ (see~\textcolor{red}{\cite{supplem_pre}}), we arrive~at}
& \Psi_{n,-m;i} = k_n(\Tilde a_i) G_{n, -m,i} + i_n(\Tilde a_i)\sum_{j=1,\, j\ne i}^N \; \sum_{L,M}\mathcal H_{n, -m}^{L M}(\mathbf R_{i j}) G_{L M,j} \notag\\
& = \left|\text{use the above symmetry}\right| = (-1)^m k_n(\Tilde a_i) G_{n m,i}^\star + i_n(\Tilde a_i) \notag\\
&\times \sum_{j=1,\, j\ne i}^N \; \sum_{L,M} (-1)^{m-M}\mathcal H_{n m}^{L, -M}(\mathbf{R}_{i j})^\star\cdot(-1)^M G_{L, -M,j}^\star \notag\\
& = \left|\text{change summation index $M\to-M$ (here $-L\le M\le L$)}\right| \notag\\
& = (-1)^m \Psi_{n,m;i}^\star \, ; \notag
\end{align}
the proof for $\Xi_{n,-m;i}$ is absolutely similar. Now using the conventional relations between local spherical basic vectors and the static Cartesian ones~\cite{Jack} we get $\mathbf n_i = \mathbf{\Hat x} \sin\theta_i \cos\varphi_i + \mathbf{\Hat y} \sin\theta_i \sin\varphi_i  + \mathbf{\Hat z} \cos\theta_i$,  $\pmb{\Hat \theta}_i = \mathbf{\Hat x} \cos\theta_i \cos\varphi_i + \mathbf{\Hat y} \cos\theta_i \sin\varphi_i  - \mathbf{\Hat z} \sin\theta_i$,  $\pmb{\Hat \varphi} = -\mathbf{\Hat x} \sin\varphi_i + \mathbf{\Hat y} \cos\varphi_i$, hence
\begin{align*}
& \Breve{\mathbf T}_{\mathbf n_i} \! = \! \Biggl[\biggl\{\!\biggl(\,\sum\limits_{n,m}\Xi_{n,m;i}Y_n^m(\Hat{\mathbf r}_i)\!\!\biggr)^{\!\! 2} - (1-\mu_i^2)\biggl(\,\sum_{n,m} \Psi_{n,m;i}\frac{\partial Y_n^m(\Hat{\mathbf r}_i)}{\partial\mu_i}\!\biggr)^{\!\! 2} \\
& - (1-\mu_i^2)^{-1}\!\biggl(\,\sum_{n,m} \Psi_{n,m;i}\frac{\partial Y_n^m(\Hat{\mathbf r}_i)}{\partial\varphi_i}\!\biggr)^{\!\! 2}\biggr\} \frac{1}{2 a_i^2} \bigl( \mathbf{\Hat x} (1-\mu_i^2)^{1/2} \cos\varphi_i \\
& + \mathbf{\Hat y} (1-\mu_i^2)^{1/2} \sin\varphi_i + \mathbf{\Hat z} \mu_i \bigr) + \biggl(\frac{-1}{a_i}\sum\limits_{n,m}\Xi_{n,m;i}Y_n^m(\Hat{\mathbf r}_i)\biggr) \\
&\times \biggl\{ \biggl(\frac{(1-\mu_i^2)^{1/2}}{a_i}\sum_{n,m} \Psi_{n,m;i}\frac{\partial Y_n^m(\Hat{\mathbf r}_i)}{\partial\mu_i}\biggr) \bigl(\mathbf{\Hat x} \mu_i \cos\varphi_i + \mathbf{\Hat y} \mu_i \sin\varphi_i \\
& - \mathbf{\Hat z} (1-\mu_i^2)^{1/2}\bigr) + \biggl(\frac{-1}{(1-\mu_i^2)^{1/2} a_i} \sum_{n,m} \Psi_{n,m;i}\frac{\partial Y_n^m(\Hat{\mathbf r}_i)}{\partial\varphi_i} \biggr) \\ 
& \times \bigl(\mathbf{\Hat y} \cos\varphi_i -\mathbf{\Hat x} \sin\varphi_i \bigr)
\biggr\} \Biggr]\varepsilon_0 \varepsilon_\text{sol}  
\intertext{and}
& \mathring{\mathbf T}_{\mathbf n_i} = -\frac{\varepsilon_0 \varepsilon_\text{sol}\kappa^2}{2} \biggl(\,\sum\limits_{n,m} \Psi_{n,m;i} Y_n^m(\Hat{\mathbf r}_i)\biggr)^{\!\! 2} \bigl( \mathbf{\Hat x} (1-\mu_i^2)^{1/2} \cos\varphi_i \\ 
&\qquad\quad + \mathbf{\Hat y} (1-\mu_i^2)^{1/2} \sin\varphi_i + \mathbf{\Hat z} \mu_i \bigr) .
\end{align*}
Now having $\Breve{\mathbf T}_{\mathbf n_i}$ and $\mathring{\mathbf T}_{\mathbf n_i}$ so expressed we perform surfacic integration \eqref{Force_i_total} separately over individual Cartesian (i.e.~$\mathbf{\Hat x}$, $\mathbf{\Hat y}$, $\mathbf{\Hat z}$) components. For instance, collecting all $\mathbf{\Hat z}$-terms in the above expression of $\Breve{\mathbf T}_{\mathbf n_i}$ we readily arrive~at
\begin{align*}
& (\Breve{\mathbf{F}}_i)_z = \oint_{\partial\Omega_i} (\Breve{\mathbf T}_{\mathbf n_i})_z \, d s = \bigl|\text{denote $D_i \mathrel{:=} \{(\varphi_i , \theta_i) \mid \varphi_i\in[0, 2\pi)$,} \\
& \qquad \text{$\theta_i\in[0, \pi]\}$} \bigr| = \biggl[ \frac{1}{2}\sum_{n,m}\sum_{k,l} \Xi_{n,m;i} \Xi_{k,l;i} \iint_{D_i} (1-\mu_i^2)^{1/2} \\
&\quad \times \mu_i Y_n^m Y_k^l \, d \varphi_i d \theta_i - \frac{1}{2} \sum_{n,m}\sum_{k,l} \Psi_{n,m;i} \Psi_{k,l;i} \iint_{D_i} (1-\mu_i^2)^{1/2} \mu_i \\
&\quad \times \biggl(\!(1-\mu_i^2)\frac{\partial Y_n^m}{\partial\mu_i}\frac{\partial Y_k^l}{\partial\mu_i} + \frac{1}{1-\mu_i^2}\frac{\partial Y_n^m}{\partial\varphi_i}\frac{\partial Y_k^l}{\partial\varphi_i} \biggr) d \varphi_i d \theta_i \\
&\quad + \sum_{n,m}\sum_{k,l} \Xi_{n,m;i} \Psi_{k,l;i} \iint_{D_i} (1-\mu_i^2)^{3/2} \, Y_n^m \frac{\partial Y_k^l}{\partial\mu_i} d \varphi_i d \theta_i \biggr]\varepsilon_0 \varepsilon_\text{sol} \\
& = \left|\text{use integrals \eqref{IntLeg1}, \eqref{IntLeg2}, \eqref{IntLeg3}}\right| \\
& = \biggl[\frac{1}{2}\sum_{n,m}\Xi_{n,m;i}(-1)^m\biggl(\sqrt{\tfrac{(n-m)(n+m)}{(2 n-1)(2 n+1)}}\Xi_{n-1,-m;i} \\
& + \sqrt{\tfrac{(n-m+1)(n+m+1)}{(2 n+1)(2 n+3)}}\Xi_{n+1,-m;i}\biggr) - \frac{1}{2}\sum_{n,m}\Psi_{n,m;i} \\
& \times (-1)^m\biggl(\sqrt{\tfrac{(n-m)(n+m)}{(2 n-1)(2 n+1)}}(n-1)(n+1)\Psi_{n-1,-m;i} \\
& + \sqrt{\tfrac{(n-m+1)(n+m+1)}{(2 n+1)(2 n+3)}} n (n+2)\Psi_{n+1,-m;i}\biggr) \\
& + \sum_{n,m}\Xi_{n,m;i}(-1)^m \biggl( \sqrt{\tfrac{(n-m)(n+m)}{(2 n-1)(2 n+1)}}(1-n)\Psi_{n-1,-m;i} \\
& + \sqrt{\tfrac{(n-m+1)(n+m+1)}{(2 n+1)(2 n+3)}} (n+2)\Psi_{n+1,-m;i}\biggr)
\biggr]\varepsilon_0 \varepsilon_\text{sol} \\
&\ = \biggl|\text{note that $\Xi_{n-1,-m;i} \equiv 0$ at $n=0$ (the corresponding} \\
&\quad  \text{integral vanishes), and we can transform the sum} \\
&\quad\  \sum_{n=1}^{+\infty}\sum_{m=-(n-1)}^{n-1} (-1)^m \Xi_{n,m;i} \Xi_{n-1,-m;i} \sqrt{\tfrac{(n-m)(n+m)}{(2 n-1)(2 n+1)}} \\ 
&\quad = \sum_{n=0}^{+\infty}\sum_{m=-n}^{n} (-1)^m \Xi_{n+1,m;i} \Xi_{n,-m;i} \sqrt{\tfrac{(n-m+1)(n+m+1)}{(2 n+1)(2 n+3)}} \\
& \quad =\left|\text{change summation index $m\to-m$}\right| \\
& \quad = \sum_{n=0}^{+\infty}\sum_{m=-n}^{n} (-1)^m \Xi_{n,m;i} \Xi_{n+1,-m;i} \sqrt{\tfrac{(n-m+1)(n+m+1)}{(2 n+1)(2 n+3)}} \, ;\\
&\quad \text{other sums can also be transformed in a similar way}\biggr| \\
& = \biggl[ \, \sum_{n,m}(-1)^m \Xi_{n,-m;i}\Xi_{n+1,m;i} \sqrt{\tfrac{(n-m+1)(n+m+1)}{(2 n+1)(2 n+3)}} \\ 
& - \sum_{n,m}(-1)^m \Psi_{n,-m;i}\Psi_{n+1,m;i} \sqrt{\tfrac{(n-m+1)(n+m+1)}{(2 n+1)(2 n+3)}} n (n+2) \\
& + \sum_{n,m}(-1)^m \bigl( (n+2)\Xi_{n,-m;i}\Psi_{n+1,m;i} - n \Xi_{n+1,m;i}\Psi_{n,-m;i} \bigr)\\
&\times \sqrt{\tfrac{(n-m+1)(n+m+1)}{(2 n+1)(2 n+3)}} \, \biggr] \varepsilon_0 \varepsilon_\text{sol} = \varepsilon_0 \varepsilon_\text{sol}\sum_{n,m} (-1)^m \\
&\times \sqrt{\tfrac{(n-m+1)(n+m+1)}{(2 n+1)(2 n+3)}} \bigl( \Xi_{n,-m;i} - n \Psi_{n,-m;i}\bigr) \\ 
& \times \bigl( \Xi_{n+1,m;i} + (n+2) \Psi_{n+1,m;i}\bigr) = \left|\text{use \eqref{Xi_Psi_for_negative_m}}\right| \\
& = \varepsilon_0 \varepsilon_\text{sol}\sum_{n,m} \sqrt{\tfrac{(n-m+1)(n+m+1)}{(2 n+1)(2 n+3)}} \bigl( \Xi_{n,m;i} - n \Psi_{n,m;i}\bigr)^\star \\ 
& \times \bigl( \Xi_{n+1,m;i} + (n+2) \Psi_{n+1,m;i}\bigr) ,
\end{align*}
which coincides with that of relations \eqref{beth_ordinary},~\eqref{force_i}. Similarly, 
\begin{align*}
& (\mathring{\mathbf{F}}_i)_z = \oint_{\partial\Omega_i} (\mathring{\mathbf T}_{\mathbf n_i})_z \, d s = 
\frac{-\varepsilon_0 \varepsilon_\text{sol}\Tilde a_i^2}{2} \sum_{n,m}\sum_{k,l} \Psi_{n,m;i} \Psi_{k,l;i} \\
&\times \iint_{D_i} (1-\mu_i^2)^{1/2} \mu_i Y_n^m Y_k^l \, d \varphi_i d \theta_i = \frac{\varepsilon_0 \varepsilon_\text{sol}\Tilde a_i^2}{-2} \sum_{n,m} \Psi_{n,m;i} (-1)^m \\
& \times \biggl(\sqrt{\tfrac{(n-m)(n+m)}{(2 n-1)(2 n+1)}}\Psi_{n-1,-m;i} + \sqrt{\tfrac{(n-m+1)(n+m+1)}{(2 n+1)(2 n+3)}}\Psi_{n+1,-m;i}\biggr) \\
& = \left|\text{perform transformations similar to those above for $(\Breve{\mathbf{F}}_i)_z$}\right| \\
& = -\varepsilon_0 \varepsilon_\text{sol}\Tilde a_i^2 \sum_{n,m}\sqrt{\tfrac{(n-m+1)(n+m+1)}{(2 n+1)(2 n+3)}}\Psi_{n+1,m;i}\Psi_{n,m;i}^\star ,
\end{align*}
which coincides with that of relations \eqref{beth_osmotic},~\eqref{force_i}. All other formulas of force components presented in \eqref{force_i}-\eqref{aleph_beth_definitions}, i.e.~$(\Breve{\mathbf{F}}_i)_x$, $(\mathring{\mathbf{F}}_i)_x$, $(\Breve{\mathbf{F}}_i)_y$, $(\mathring{\mathbf{F}}_i)_y$, can be derived in a similar way. 

Finally, taking into account that for any real-valued function $f$ the coincidence of representations $f=\sum\limits_{n=0}^{+\infty}\sum\limits_{m=0}^n (M^c_{n m}\cos(m\varphi) + M^s_{n m}\sin(m\varphi)) P_n^m(\cos\theta) = \sum\limits_{n=0}^{+\infty}\sum\limits_{m=-n}^n M_{n m} Y_n^m(\theta,\varphi)$ is equivalent to obeying coefficients equalities 
\begin{subequations}
\label{eqs_Ynm_Pnm_}
\begin{align}
M^c_{n 0} & = \sqrt{\tfrac{2 n+1}{4\pi}} M_{n 0} && (n\ge0), \label{eqs_Ynm_Pnm_1}\\
M^c_{n m} & = \sqrt{\tfrac{(2 n+1) (n-m)!}{\pi (n+m)!}}\operatorname{Re} M_{n m} && (n\ge m\ge 1),\label{eqs_Ynm_Pnm_2}\\
M^s_{n m} &= -\sqrt{\tfrac{(2 n+1) (n-m)!}{\pi (n+m)!}}\operatorname{Im} M_{n m} && (n\ge m\ge 1), \label{eqs_Ynm_Pnm_3}
\end{align}
\end{subequations}
the equivalence between \eqref{force_i}-\eqref{aleph_beth_definitions} and \eqref{force_Leg_} can then be established after bulky algebraic transformations.
\begin{remark}
\label{various_Leg_integrals_remark}
Let us finally note that when deriving the above formulas for the force components, one faces the need to analytically calculate integrals of various combinations of spherical harmonics and their derivatives; while integration over $\varphi_i$ can be accomplished trivially (and is advisable to be performed first, in general), e.g.~in the above calculations of $z$-components we simply have $\int_0^{2\pi} e^{\imath (m+l) \varphi_i} d \varphi_i = 2\pi\delta_{l,-m}$, integration over $\theta_i$ is more technical -- in~\eqref{IntsLeg} we collected a complete list of the corresponding integrals necessary to derive the formulas proposed in Sec.~\ref{explicit_forces_section}.
\end{remark}

\section{Non-screened (``0-screened") components of interaction force are zero: a rigorous proof}
\label{non-screened-force-components-vanish-appendix}
\noindent
Indeed, from the very definitions of $\aleph_i^{(\ell)}$ and $\beth_i^{(\ell)}$ (see \cite{supplem_prl}) we have 
$$
 \Psi_{n,m;i}^{(0)} = k_n(\Tilde a_i) G_{n m,i}^{(0)}, \quad \Xi_{n,m;i}^{(0)} = \bigl(n k_n(\Tilde a_i) - \Tilde a_i k_{n+1}(\Tilde a_i)\bigr) G_{n m,i}^{(0)} ,
$$ 
thus
\begin{gather*}
 A_{n,m;i}^{(0)} = -\Tilde a_i k_{n+1}(\Tilde a_i) G_{n m,i}^{(0)}, \\
 B_{n,m;i}^{(0)} = \bigl((2 n+1) k_n(\Tilde a_i) - \Tilde a_i k_{n+1}(\Tilde a_i)\bigr) G_{n m,i}^{(0)},
\end{gather*}
from which we then obtain
\begin{align*}
\aleph_i^{(0)} & = \varepsilon_0\varepsilon_\text{sol}\sum_{n,m}\sqrt{\tfrac{(n-m+1) (n-m+2)}{(2 n+1) (2 n+3)}} \bigl( B_{n+1,m-1;i}^{(0)} A_{n,m;i}^{(0)}{}^\star \\ 
& - \Tilde a_i^2 \Psi_{n+1,m-1;i}^{(0)}\Psi_{n,m;i}^{(0)}{}^\star\bigr) = \varepsilon_0\varepsilon_\text{sol}\sum_{n,m}\sqrt{\tfrac{(n-m+1) (n-m+2)}{(2 n+1) (2 n+3)}} \\ &\times \Bigl(\!\bigl((2 n+3) k_{n+1}(\Tilde a_i) - \Tilde a_i k_{n+2}(\Tilde a_i)\bigr) \bigl(-\Tilde a_i k_{n+1}(\Tilde a_i) \bigr) \\ 
& - \Tilde a_i^2 k_{n+1}(\Tilde a_i) k_n(\Tilde a_i)\!\Bigr)G_{n+1,m-1,i}^{(0)} G_{n m,i}^{(0)}{}^\star \\
& = \varepsilon_0\varepsilon_\text{sol}\sum_{n,m}\sqrt{\tfrac{(n-m+1) (n-m+2)}{(2 n+1) (2 n+3)}} \Bigl(\! \bigl( k_{n+2}(\Tilde a_i)-k_{n}(\Tilde a_i)\bigr)\Tilde a_i \\ 
& - (2 n+3) k_{n+1}(\Tilde a_i) \!\Bigr) \Tilde a_i k_{n+1}(\Tilde a_i) G_{n+1,m-1,i}^{(0)} G_{n m,i}^{(0)}{}^\star \\
& = \Bigl|\text{use recurrence \eqref{in_kn_recurrences0} for modified Bessel functions:} \\ 
& \bigl(k_{n+2}(\Tilde a_i)-k_n(\Tilde a_i)\bigr)\Tilde a_i - (2 n+3) k_{n+1}(\Tilde a_i) = 0\Bigr| \\
& = 0
\end{align*}
(the proof of $\beth_i^{(0)}=0$ is completely similar, but with $G_{n+1,m-1,i}^{(0)}$ replaced by $G_{n+1,m,i}^{(0)}$).

\section{Derivation of~\eqref{force_kappa_zero_BBKS_0}}
\label{force_kappa_zero_BBKS_Appendix_derivation}
\noindent
From \eqref{force_i}-\eqref{aleph_beth_definitions} we obtain in this azimuthally symmetric case: $(\mathbf F_1)_x=(\mathbf F_1)_y=0$ and
\begin{equation}
\label{F_z_twospheres_kappa0_bb}
(\mathbf F_1)_z = \varepsilon_0\varepsilon_\text{sol}\sum_{n=0}^{+\infty}\frac{n+1}{\sqrt{(2 n+1)(2 n+3)}} B_{n+1,0;1} A_{n,0;1}^\star .
\end{equation}

In order to investigate the behavior of expression \eqref{F_z_twospheres_kappa0_bb} as $\kappa\to0$ we will need relation~\textcolor{red}{\cite{supplem_pre}}
\begin{equation}
\label{b_H_equality_}
\begin{aligned}
& \sqrt{2/\pi}\sum\nolimits_{l=0}^{+\infty} \sqrt{\tfrac{2 l+1}{2 n+1}} b_{n 0 l}(\Tilde r_1,\Tilde R) (-1)^l G_{l 0,2} \\
& \quad = i_n(\Tilde r_1)\sum\nolimits_{L=0}^{+\infty} \mathcal H_{n 0}^{L 0}(R \Hat{\mathbf z}) G_{L 0,2} ,
\end{aligned}
\end{equation}
where a detailed definition of the quantities $b_{n 0 l}$ is given in \cite[Eq.~(37)]{our_jcp}, as well as the asymptotics of the latter \cite[Eq.~(B7)]{our_jcp} as $\kappa\to0$:
\begin{equation}
\label{b_n0l_asympt0_}
b_{n 0 l}(\Tilde r, \Tilde R) \sim \sqrt{\frac{\pi}{2}}(2 l-1)!!\frac{(n+l)!}{n! l!}\frac{r^n}{R^{n+l+1}\kappa^{l+1}} .
\end{equation}

Next, in the $\kappa\to0$ limit, external potentials $\Phi_{\text{out},i}$ satisfy the Laplace equation $\Delta\Phi_{\text{out},i}=0$, thus taking into account the azimuthal symmetry of the problem under consideration one gets $\Phi_{\text{out},1}(\mathbf r) = \sum_{l=0}^{+\infty}\mathfrak G_{l,1} r_1^{-l-1} P_l(\cos\theta_1)$ and $\Phi_{\text{out},2}(\mathbf r) = \sum_{l=0}^{+\infty}\mathfrak G_{l,2} r_2^{-l-1} P_l(\cos\theta_2')$, where $\theta_2'=\pi-\theta_2$ is the complementary angle to the polar angle $\theta_2$ used throughout this study by default (here we make use $\theta_2'$ instead of $\theta_2$ to obtain the relations compatible with the notations of work \cite{BBKS} as well as to employ below the two-center addition theorem for solid harmonics, where such notations are conventional). Comparing these expansions with \eqref{Lin_eqs_Phi_out} at finite $\kappa$ and using \eqref{small_Bessel_i_k} and \eqref{Ynm_definition} one may anticipate that $\mathfrak G_{l,1}\sim G_{l 0,1}\frac{(2 l-1)!!}{\kappa^{l+1}}\sqrt{\frac{2 l+1}{4\pi}}$ and $\mathfrak G_{l,2}\sim G_{l 0,2}\frac{(2 l-1)!!}{\kappa^{l+1}}\sqrt{\frac{2 l+1}{4\pi}}(-1)^l$ as $\kappa\to0$ (see also \cite[Appendix~F]{our_jpcb}). Now using these, relations \eqref{b_H_equality_}, \eqref{b_n0l_asympt0_}, \eqref{in_kn_recurrences0}, and asymptotics \eqref{small_Bessel_i_k}, we arrive at the following results after algebraic transformations:
\begin{align*}
& B_{n,0;1}\xrightarrow[\kappa\to0]{}\sqrt{2n+1}\sum_{l=0}^{+\infty}\sqrt{4\pi}\frac{(l+n)!}{l! n!}\frac{a_1^n \mathfrak G_{l,2}}{R^{n+l+1}} ,\\
& A_{n,0;1}\xrightarrow[\kappa\to0]{}-\sqrt{4\pi(2 n+1)}\mathfrak G_{n,1}/a_1^{n+1} ,
\end{align*}
employing which in \eqref{F_z_twospheres_kappa0_bb} we finally obtain the following relation for the force in the $\kappa\to0$ limit:
\begin{equation}
\label{F_z_twospheres_kappa0_bb_1}
(\mathbf F_1)_z = -4\pi\varepsilon_0\varepsilon_\text{sol}\sum_{n=0}^{+\infty}\sum_{l=0}^{+\infty}\frac{(n+l+1)!}{n! l! R^{n+l+2}} \mathfrak G_{n,1} \mathfrak G_{l,2} ,
\end{equation}
which recovers \cite[Eq.~(26)]{BBKS}. 

Finally, the desired relations \eqref{force_kappa_zero_BBKS_0} easily result from \eqref{F_z_twospheres_kappa0_bb_1} by combining it with the following relation \cite{BBKS, LCSB} readily stemming from boundary conditions~\eqref{Lin_eq_standard_bc}:
\begin{equation}
\label{Lin_eq_standard_bc_kappa_0_consequen}
\mathfrak G_{n,i} + \frac{n (k_i-1) a_i^{2 n+1}}{n(k_i+1)+1}\sum_{l=0}^{+\infty}\frac{(l+n)! \mathfrak G_{l,j}}{n! l! R^{n+l+1}} = \frac{q_i \delta_{n,0}}{4\pi\varepsilon_0\varepsilon_\text{sol}} 
\end{equation}
with $i=\overline{1, 2}$ and $j=3-i$. (In order to obtain relation \eqref{Lin_eq_standard_bc_kappa_0_consequen} one can represent internal potentials as $\Phi_{\text{in},1}(\mathbf r) = \sum_{l=0}^{+\infty}\mathfrak L_{l,1} r_1^l P_l(\cos\theta_1)$ and $\Phi_{\text{in},2}(\mathbf r) = \sum_{l=0}^{+\infty}\mathfrak L_{l,2} r_2^l P_l(\cos\theta_2')$ and then use the two-center addition theorem for solid harmonics, see e.g.~\cite[Appendix~F]{our_jpcb}.)

\section{Details on calculation of forces in a system of $N$ spheres with centrally-located point charges}
\label{Appendix_screened_potential_coefficients_point_force}
\noindent
In this particular case, applying the explicit expressions for $\{\Tilde G_{n m,i}^{(\ell)}\}$ (see \textcolor{red}{\cite{supplem_pre}}) and using the relations $i_n(\Tilde a_i) \mathcal H_{n m}^{L M}(\mathbf R_{i j}) - \frac{k_n(\Tilde a_i)}{\alpha_n(\Tilde a_i,\varepsilon_i)}\beta_{n m, L M}(\Tilde a_i,\varepsilon_i,\mathbf R_{i j}) = \frac{\varepsilon_\text{sol}}{\Tilde a_i^2 \alpha_n(\Tilde a_i,\varepsilon_i)}\mathcal H_{n m}^{L M}(\mathbf R_{i j})$ and $\left(n i_n(\Tilde a_i) + \Tilde a_i i_{n+1}(\Tilde a_i)\right)\mathcal H_{n m}^{L M}(\mathbf R_{i j}) - \frac{n k_n(\Tilde a_i) - \Tilde a_i k_{n+1}(\Tilde a_i)}{\alpha_n(\Tilde a_i,\varepsilon_i)} \beta_{n m, L M}(\Tilde a_i,\varepsilon_i,\mathbf R_{i j}) = \frac{n\varepsilon_i}{\Tilde a_i^2 \alpha_n(\Tilde a_i,\varepsilon_i)}\mathcal H_{n m}^{L M}(\mathbf R_{i j})$ established in~\textcolor{red}{\cite{supplem_pre}}, we readily obtain the corresponding screening-ranged expressions of quantities $\Xi_{n,m;i}^{(\ell)}$, $\Psi_{n,m;i}^{(\ell)}$, $A_{n,m;i}^{(\ell)}$, $B_{n,m;i}^{(\ell)}$ appearing in the general screening-ranged force expansion~\eqref{Force_expansion_vec}:
\begin{align*}
\Psi_{n,m;i}^{(0)} & = \frac{(2 n+1)k_n(\Tilde a_i) q_i \delta_{n,0} \kappa}{\sqrt{4\pi}\varepsilon_0\Tilde a_i^{n+2}\alpha_n(\Tilde a_i,\varepsilon_i)} ,\\
\Xi_{n,m;i}^{(0)} & = \frac{(n k_n(\Tilde a_i)-\Tilde a_i k_{n+1}(\Tilde a_i))(2 n+1) q_i \delta_{n,0} \kappa}{\sqrt{4\pi}\varepsilon_0\Tilde a_i^{n+2}\alpha_n(\Tilde a_i,\varepsilon_i)} ,\\
A_{n,m;i}^{(0)} & = \frac{-q_i\kappa\delta_{n,0}}{\sqrt{4\pi}\varepsilon_0\varepsilon_\text{sol}\Tilde a_i}, \qquad B_{n,m;i}^{(0)} = \frac{-q_i\kappa\delta_{n,0}}{\sqrt{4\pi}\varepsilon_0\varepsilon_\text{sol}(1+\Tilde a_i)}, \\
\Psi_{n,m;i}^{(1)} & = \frac{\kappa}{\varepsilon_0\Tilde a_i^2\alpha_n(\Tilde a_i,\varepsilon_i)}\sum_{j=1,\, j\ne i}^N \frac{q_j e^{\Tilde a_j}}{1+\Tilde a_j}k_n(\Tilde R_{i j}) Y_n^m(\Hat{\mathbf R}_{i j})^\star , \\ 
\Xi_{n,m;i}^{(1)} & = \frac{n \kappa \varepsilon_i}{\varepsilon_0\varepsilon_\text{sol}\Tilde a_i^2\alpha_n(\Tilde a_i,\varepsilon_i)}\sum_{j=1,\, j\ne i}^N \frac{q_j e^{\Tilde a_j}}{1+\Tilde a_j}k_n(\Tilde R_{i j}) Y_n^m(\Hat{\mathbf R}_{i j})^\star , \\
A_{n,m;i}^{(1)} & = \frac{(\varepsilon_i-\varepsilon_\text{sol}) n \kappa}{\varepsilon_0 \varepsilon_\text{sol} \Tilde a_i^2 \alpha_n(\Tilde a_i,\varepsilon_i)}\sum_{j=1,\, j\ne i}^N \frac{q_j e^{\Tilde a_j}}{1+\Tilde a_j}k_n(\Tilde R_{i j}) Y_n^m(\Hat{\mathbf R}_{i j})^\star , \\
B_{n,m;i}^{(1)} & = \frac{(n\varepsilon_i+(n+1)\varepsilon_\text{sol}) \kappa}{\varepsilon_0 \varepsilon_\text{sol} \Tilde a_i^2 \alpha_n(\Tilde a_i,\varepsilon_i)}\sum_{j=1,\, j\ne i}^N \frac{q_j e^{\Tilde a_j}}{1+\Tilde a_j}k_n(\Tilde R_{i j}) Y_n^m(\Hat{\mathbf R}_{i j})^\star , \\
\Psi_{n,m;i}^{(2)} & = \frac{-\kappa}{\sqrt{4\pi}\varepsilon_0\Tilde a_i^2\alpha_n(\Tilde a_i,\varepsilon_i)} \sum_{j=1,\, j\ne i}^N \; \sum_{L, M} \frac{\mathcal H_{n m}^{L M}(\mathbf{R}_{i j})}{\alpha_L(\Tilde a_j,\varepsilon_j)} \\ 
&\quad\times\sum_{p=1,\, p\ne j}^N \frac{q_p e^{\Tilde a_p}}{1+\Tilde a_p} \beta_{L M, 0 0}(\Tilde a_j,\varepsilon_j,\mathbf R_{j p}) , \\
\Xi_{n,m;i}^{(2)} & = \frac{-n\kappa\varepsilon_i}{\sqrt{4\pi}\varepsilon_0\varepsilon_\text{sol}\Tilde a_i^2\alpha_n(\Tilde a_i,\varepsilon_i)} \sum_{j=1,\, j\ne i}^N \; \sum_{L, M} \frac{\mathcal H_{n m}^{L M}(\mathbf{R}_{i j})}{\alpha_L(\Tilde a_j,\varepsilon_j)} \\ 
&\quad\times\sum_{p=1,\, p\ne j}^N \frac{q_p e^{\Tilde a_p}}{1+\Tilde a_p} \beta_{L M, 0 0}(\Tilde a_j,\varepsilon_j,\mathbf R_{j p}) , \\
A_{n,m;i}^{(2)} & = \frac{(\varepsilon_\text{sol}-\varepsilon_i) n \kappa}{\sqrt{4\pi}\varepsilon_0\varepsilon_\text{sol}\Tilde a_i^2\alpha_n(\Tilde a_i,\varepsilon_i)} \sum_{j=1,\, j\ne i}^N \; \sum_{L, M} \frac{\mathcal H_{n m}^{L M}(\mathbf{R}_{i j})}{\alpha_L(\Tilde a_j,\varepsilon_j)} \\ 
&\quad\times\sum_{p=1,\, p\ne j}^N \frac{q_p e^{\Tilde a_p}}{1+\Tilde a_p} \beta_{L M, 0 0}(\Tilde a_j,\varepsilon_j,\mathbf R_{j p}) , \\
B_{n,m;i}^{(2)} & = \frac{-(n\varepsilon_i+(n+1)\varepsilon_\text{sol}) \kappa}{\sqrt{4\pi}\varepsilon_0\varepsilon_\text{sol}\Tilde a_i^2\alpha_n(\Tilde a_i,\varepsilon_i)} \sum_{j=1,\, j\ne i}^N \; \sum_{L, M} \frac{\mathcal H_{n m}^{L M}(\mathbf{R}_{i j})}{\alpha_L(\Tilde a_j,\varepsilon_j)} \\ 
&\quad\times\sum_{p=1,\, p\ne j}^N \frac{q_p e^{\Tilde a_p}}{1+\Tilde a_p} \beta_{L M, 0 0}(\Tilde a_j,\varepsilon_j,\mathbf R_{j p}) .
\end{align*}
From these expressions (and taking into account identities for low-order $\mathcal H_{n m}^{L M}(\mathbf R_{i j})$ established in \textcolor{red}{\cite{supplem_pre}}) we can now calculate $\Breve\aleph_i^{(\ell)}$, $\mathring\aleph_i^{(\ell)}$, $\Breve\beth_i^{(\ell)}$, $\mathring\beth_i^{(\ell)}$, $\ell=\overline{1, 2}$ -- e.g.~according to our general formulas (see Sec.~\ref{general_expansions_pot_force}) we have $\Breve\aleph_i^{(2)} = \varepsilon_0\varepsilon_\text{sol}\sum\limits_{n,m} \sqrt{\frac{(n-m+1) (n-m+2)}{(2 n+1) (2 n+3)}} \bigl( B_{n+1,m-1;i}^{(0)} A_{n,m;i}^{(2)}{}^\star + B_{n+1,m-1;i}^{(1)} A_{n,m;i}^{(1)}{}^\star + B_{n+1,m-1;i}^{(2)} A_{n,m;i}^{(0)}{}^\star \bigr)$, $\Breve\beth_i^{(2)} = \varepsilon_0\varepsilon_\text{sol}\sum\limits_{n,m} \sqrt{\frac{(n-m+1) (n+m+1)}{(2 n+1) (2 n+3)}} \bigl( B_{n+1,m;i}^{(0)} A_{n,m;i}^{(2)}{}^\star + B_{n+1,m;i}^{(1)} A_{n,m;i}^{(1)}{}^\star + B_{n+1,m;i}^{(2)} A_{n,m;i}^{(0)}{}^\star \bigr)$, and so on -- their expanded forms up to the 1st-order addends ($n,L\le1$) read~as
\begin{widetext}
\begin{subequations}
\label{monopoles_force_double_all}
\begin{align}
&\!\!\!\!\begin{aligned}
& \!\!\! \Breve\aleph_i^{(2)} = \frac{d_{3,i} d_{4,i} \kappa^2}{8\pi\varepsilon_0\varepsilon_\text{sol}}\!\!\sum_{j=1, j\ne i}^N \; \sum_{p=1, p\ne i}^N \frac{q_j q_p e^{\Tilde a_j + \Tilde a_p} e^{-\Tilde R_{i j}} e^{-\Tilde R_{i p}}}{(1+\Tilde a_j) (1+\Tilde a_p) \Tilde R_{i j} \Tilde R_{i p}}\!\Bigl(\! 1 \!+\! \frac{1}{\Tilde R_{i p}} \Bigr) \Lambda_{i j i p}^\aleph  -\frac{q_i e^{\Tilde a_i} d_{1,i} \kappa^2}{8\pi\varepsilon_0\varepsilon_\text{sol}}\!\!\sum_{j=1, j\ne i}^N \; \sum_{p=1, p\ne j}^N \frac{q_p e^{\Tilde a_p}}{1+\Tilde a_p}\\
&\!\!\! \times \frac{e^{-\Tilde R_{i j}}}{\Tilde R_{i j}}\!\frac{e^{-\Tilde R_{j p}}}{\Tilde R_{j p}}\biggl[\Bigl(1+\frac{1}{\Tilde R_{i j}}\Bigr) d_{0,j} \sin\theta_{i j} e^{\imath\varphi_{i j}} + \biggl\{ \sin\theta_{j p} e^{\imath\varphi_{j p}} +  \frac{1}{2}\Lambda_{i j j p}^\aleph \biggr\}\Bigl(1+\frac{1}{\Tilde R_{j p}}\Bigr) d_{2,j} \biggr] + \text{H.O.T.} ,
\end{aligned} \label{monopoles_xy_double_standard} \\
&\!\!\!\!\begin{aligned}
& \!\!\! \mathring\aleph_i^{(2)} = \frac{-\kappa^2\Tilde a_i^2}{4\pi\varepsilon_0} \!\!\sum_{j=1, j\ne i}^N \; \sum_{p=1, p\ne i}^N \frac{q_j q_p e^{\Tilde a_j + \Tilde a_p} e^{-\Tilde R_{i j}} e^{-\Tilde R_{i p}}}{(1+\Tilde a_j) (1+\Tilde a_p) \Tilde R_{i j} \Tilde R_{i p}}\! \biggl[ \frac{e^{2\Tilde a_i}\Tilde a_i d_{1,i}^\prime}{1+\Tilde a_i}\Bigl(1+\frac{1}{\Tilde R_{i j}}\Bigr) \sin\theta_{i j}e^{\imath\varphi_{i j}} + \frac{\varepsilon_\text{sol}d_{3,i}^\prime d_{4,i}^\prime}{2}\!\Bigl(\!1\!+\!\frac{1}{\Tilde R_{i p}}\!\Bigr) \Lambda_{i j i p}^\aleph \biggr] \\
&\!\!\! - \frac{q_i e^{\Tilde a_i} d_{1,i}^\prime \kappa^2 \Tilde a_i^2}{8\pi\varepsilon_0(1+\Tilde a_i)}\!\!\sum_{j=1, j\ne i}^N \: \sum_{p=1, p\ne j}^N \!\frac{q_p e^{\Tilde a_p}}{1+\Tilde a_p}\! \frac{e^{-\Tilde R_{i j}}}{\Tilde R_{i j}}\!\frac{e^{-\Tilde R_{j p}}}{\Tilde R_{j p}}\!\!\biggl[\!\Bigl(\!1\!+\!\frac{1}{\Tilde R_{i j}}\!\Bigr) d_{0,j} \sin\theta_{i j} e^{\imath\varphi_{i j}} \!+\! \biggl\{\! \sin\theta_{j p} e^{\imath\varphi_{j p}} \!+\!  \frac{\Lambda_{i j j p}^\aleph}{2}\! \biggr\}\!\Bigl(\!1\!+\!\frac{1}{\Tilde R_{j p}}\!\Bigr) d_{2,j} \!\biggr]\! + \!\text{H.O.T.} , 
\end{aligned} \label{monopoles_xy_double_osmotic} \\
&\!\!\!\!\begin{aligned}
& \!\!\! \Breve\beth_i^{(2)} = \frac{d_{3,i} d_{4,i} \kappa^2}{4\pi\varepsilon_0\varepsilon_\text{sol}}\!\!\sum_{j=1, j\ne i}^N \; \sum_{p=1, p\ne i}^N \frac{q_j q_p e^{\Tilde a_j + \Tilde a_p} e^{-\Tilde R_{i j}} e^{-\Tilde R_{i p}}}{(1+\Tilde a_j) (1+\Tilde a_p) \Tilde R_{i j} \Tilde R_{i p}}\!\Bigl(\! 1 \!+\! \frac{1}{\Tilde R_{i p}} \Bigr)\Lambda_{i j i p}^\beth  -\frac{q_i e^{\Tilde a_i} d_{1,i} \kappa^2}{8\pi\varepsilon_0\varepsilon_\text{sol}}\!\!\sum_{j=1, j\ne i}^N \; \sum_{p=1, p\ne j}^N \frac{q_p e^{\Tilde a_p}}{1+\Tilde a_p}\\
&\!\!\! \times \frac{e^{-\Tilde R_{i j}}}{\Tilde R_{i j}}\!\frac{e^{-\Tilde R_{j p}}}{\Tilde R_{j p}}\biggl[\Bigl(1+\frac{1}{\Tilde R_{i j}}\Bigr) d_{0,j} \cos\theta_{i j} + \Bigl\{ \cos\theta_{j p} +  \Lambda_{i j j p}^\beth \Bigr\}\Bigl(1+\frac{1}{\Tilde R_{j p}}\Bigr) d_{2,j} \biggr] + \text{H.O.T.},
\end{aligned} \label{monopoles_z_double_standard} \\
&\!\!\!\!\begin{aligned}
& \!\!\! \mathring\beth_i^{(2)} = \frac{-\kappa^2\Tilde a_i^2}{4\pi\varepsilon_0} \!\!\sum_{j=1, j\ne i}^N \; \sum_{p=1, p\ne i}^N \frac{q_j q_p e^{\Tilde a_j + \Tilde a_p} e^{-\Tilde R_{i j}} e^{-\Tilde R_{i p}}}{(1+\Tilde a_j) (1+\Tilde a_p) \Tilde R_{i j} \Tilde R_{i p}}\! \biggl[ \frac{e^{2\Tilde a_i}\Tilde a_i d_{1,i}^\prime}{1+\Tilde a_i}\Bigl(1+\frac{1}{\Tilde R_{i j}}\Bigr) \cos\theta_{i j} + \varepsilon_\text{sol}d_{3,i}^\prime d_{4,i}^\prime\!\Bigl(\!1\!+\!\frac{1}{\Tilde R_{i p}}\!\Bigr) \Lambda_{i j i p}^\beth \biggr] \\
&\!\!\! - \frac{q_i e^{\Tilde a_i} d_{1,i}^\prime \kappa^2 \Tilde a_i^2}{8\pi\varepsilon_0(1+\Tilde a_i)}\!\!\sum_{j=1, j\ne i}^N \; \sum_{p=1, p\ne j}^N \frac{q_p e^{\Tilde a_p}}{1+\Tilde a_p} \frac{e^{-\Tilde R_{i j}}}{\Tilde R_{i j}}\!\frac{e^{-\Tilde R_{j p}}}{\Tilde R_{j p}}\!\!\biggl[\!\Bigl(\!1\!+\!\frac{1}{\Tilde R_{i j}}\!\Bigr) d_{0,j} \cos\theta_{i j} + \Bigl\{\! \cos\theta_{j p} \!+\! \Lambda_{i j j p}^\beth\! \Bigr\}\!\Bigl(\!1\!+\!\frac{1}{\Tilde R_{j p}}\!\Bigr) d_{2,j} \!\biggr] + \text{H.O.T.} , 
\end{aligned} \label{monopoles_z_double_osmotic}
\end{align} 
\end{subequations}
where $\theta_{i j}$ and $\varphi_{i j}$ are spherical angles of $\Hat{\mathbf R}_{i j}$, $\Lambda_{i_1 i_2 i_3 i_4}^\aleph \mathrel{:=} \bigl( 3 \sin^2\theta_{i_1 i_2} \sin\theta_{i_3 i_4} e^{\imath(2\varphi_{i_1 i_2}-\varphi_{i_3 i_4})} + 3\sin(2\theta_{i_1 i_2})\cos\theta_{i_3 i_4} e^{\imath\varphi_{i_1 i_2}} + (1-3\cos^2\theta_{i_1 i_2}) \sin\theta_{i_3 i_4} e^{\imath\varphi_{i_3 i_4}} \bigr) \Bigl(1+\frac{3}{\Tilde R_{i_1 i_2}}+\frac{3}{\Tilde R_{i_1 i_2}^2}\Bigr)$, $\Lambda_{i_1 i_2 i_3 i_4}^\beth \mathrel{:=} \bigl( 3\sin\theta_{i_1 i_2}\cos\theta_{i_1 i_2}\sin\theta_{i_3 i_4}\cos(\varphi_{i_1 i_2}-\varphi_{i_3 i_4})+(3\cos^2\theta_{i_1 i_2}-1)\cos\theta_{i_3 i_4}\bigr)\Bigl(1+\frac{3}{\Tilde R_{i_1 i_2}}+\frac{3}{\Tilde R_{i_1 i_2}^2}\Bigr)$, $\imath$ is a complex unit, and auxiliary symbols $d_{0,i}$, $d_{2,i}$, $d_{1,i}^\prime$, $d_{3,i}^\prime$, $d_{4,i}^\prime$, $d_{1,i}$, $d_{3,i}$, $d_{4,i}$ were defined in Sec.~\ref{Appendix_screened_potential_coefficients_point_force_section}. Relations \eqref{monopoles_force_double_all} then yield~\eqref{forces_double_screening_all_} after trivial trigonometric transformations of the above angular factors $\Lambda_{i_1 i_2 i_3 i_4}^\aleph$,~$\Lambda_{i_1 i_2 i_3 i_4}^\beth$.
\end{widetext}

\section{Equalities for determining the addends in~\eqref{G_componentwise0_}}
\label{G_appendix_nm_ell_elementwise}
\noindent
Introducing the block matrix $\mathbb K \mathrel{:=} \{\mathsf A_i^{-1} \mathsf B_{i j}\}_{i,j=1;\, i\ne j}^N$ consisting of subblocks $\mathsf A_i^{-1} \mathsf B_{i j}$ (see \eqref{various_matrix_definitions_0}), then it can be shown \textcolor{red}{\cite{supplem_pre}} that the addends of \eqref{G_componentwise0_} are determined by $\Tilde{\mathbf G}_i^{(\ell)} = (-1)^\ell\sum_{j=1}^N (\mathbb K^\ell)_{i j} \mathsf A_j^{-1}\mathbf S_j$ -- or, in expanded form:
\begin{align*}
& \Tilde{\mathbf G}_i^{(0)} = \mathsf A_i^{-1}\mathbf S_i , \\ 
& \Tilde{\mathbf G}_i^{(1)} = -\mathsf A_i^{-1}\sum\nolimits_{j=1,\, j\ne i}^N\mathsf B_{i j} \mathsf A_j^{-1}\mathbf S_j , \\
& \Tilde{\mathbf G}_i^{(2)} = \mathsf A_i^{-1}\sum\nolimits_{j=1}^N\sum\nolimits_{k=1,\, k\ne i,\, k\ne j}^N\mathsf B_{i k} \mathsf A_k^{-1}\mathsf B_{k j} \mathsf A_j^{-1} \mathbf S_j , \\ 
& \ldots \ .
\end{align*}
Here matrix $(\mathbb K^\ell)_{i j}$ is the $(i,j)$-th block of block matrix $\mathbb K^\ell$, the $\ell$-th power of $\mathbb K$. The elements of the general vector $\Tilde{\mathbf G}_i^{(\ell)}$ for $\forall\ell\ge0$, $0\le |m|\le n$, $1\le i\le N$~are 
$$
\Tilde G^{(\ell)}_{n m,i} = (-1)^\ell\sum_{j=1}^N \, \sum_{L,M}\bigl((\mathbb K^\ell)_{i j}\bigr)_{n m, L M}\frac{S_{L M,j}}{\alpha_L(\Tilde a_j,\varepsilon_j) \Upsilon_{L,j}} ,
$$
where $S_{n m,i} = (2 n+1)\varepsilon_i\Hat L_{n m,i}/\Tilde a_i^{n+2}$ are elements of vector $\mathbf S_i$ (see~\eqref{various_matrix_definitions_0}).

\section{Selected facts on modified Bessel functions and spherical~harmonics}
\label{appendix_bessel_functions_summary}
\noindent
Spherical modified Bessel functions of the 1st and 2nd kinds, respectively $i_n(x)$ and $k_n(x)$, are defined via modified Bessel functions of the 1st and 2nd kinds of semi-integer order $n+1/2$, respectively $I_{n+1/2} ( \cdot )$ and $K_{n+1/2} ( \cdot )$, as $i_n(x) \mathrel{:=} \sqrt{\frac{\pi}{2}}\frac{I_{n+1/2}(x)}{\sqrt{x}}$ and $k_n(x) \mathrel{:=} \sqrt{\frac{2}{\pi}}\frac{K_{n+1/2}(x)}{\sqrt{x}}$.

There are recurrences~\cite[Eq.~(8.486)]{GradRyzh} 
\begin{equation}
\label{in_kn_recurrences0}
\begin{aligned}
& x k_{n+1}(x) = x k_{n-1}(x) + (2 n+1)k_n(x), \\ 
& x i_{n+1}(x) = x i_{n-1}(x) - (2 n+1)i_n(x).
\end{aligned}
\end{equation}
In addition, for their derivatives there are the expressions
\begin{equation}
\label{diff_modifiedBessel}
i'_n(x) = \frac{n}{x} i_n(x) + i_{n+1}(x), \quad  
k'_n(x) = \frac{n}{x} k_n(x) - k_{n+1}(x).
\end{equation}

For small $x\to0^+$ one has~\cite{Wat}
\begin{equation}
\label{small_Bessel_i_k}
\begin{aligned}
i_n(x) & = \frac{x^n}{(2 n+1)!!} + \frac{x^{n+2}}{2 (2 n+3)!!} + O(x^{n+4}) \\ 
& \sim \frac{x^n}{(2 n+1)!!} \quad \text{and} \quad  k_n(x) \sim \frac{(2 n-1)!!}{x^{n+1}},
\end{aligned}
\end{equation}
while for large $x\to+\infty$ one has $k_n(x) \sim \frac{e^{-x}}{x}$ (notation ``$f(x)\sim g(x)$ as $x\to y$'' here means that $f(x)$ behaves asymptotically like $g(x)$ as $x\to y$, $\lim_{x\to y}(f(x)/g(x))=1$). 

Spherical (complex-valued) harmonics are defined as~\cite{Jack} 
\begin{equation}
\label{Ynm_definition}
Y_n^m(\Hat{\mathbf r}_i) = \sqrt{\tfrac{(2 n+1)(n-m)!}{4\pi(n+m)!}} P_n^m(\cos\theta_i) e^{\imath m\varphi_i} ,
\end{equation}
where $P_n^m(x)$ denotes the conventional associated Legendre polynomial, $P_n^m(x) = \frac{(-1)^m}{2^n n!}(1-x^2)^{m/2}\frac{d^{n+m}}{d x^{n+m}}(x^2-1)^n$, and $\imath$ is a complex unit. When deriving explicit formulas for forces (especially in Sec.~\ref{explicit_forces_section}, Appendix~\ref{appendix_force_derivation}) we will need to integrate the harmonics over variable $\mu_i\mathrel{:=}\cos\theta_i$; a complete list of the particularly useful integrals we will need is the following: \begin{widetext}
\begin{subequations}
\label{IntsLeg}
\begin{align}
& \int_{-1}^1 P_n^m(\mu_i) P_k^m(\mu_i) \mu_i \, d\mu_i = \frac{2}{2 n+1} \biggl(\frac{\delta_{k,n-1} (n+m)!}{(2 n-1) (n-m-1)!} + \frac{\delta_{k,n+1} (n+m+1)!}{(2 n+3) (n-m)!}  \biggr), \label{IntLeg1}\\
& \int_{-1}^1 (1-\mu_i^2) P_n^m(\mu_i) \frac{d P_k^m(\mu_i)}{d \mu_i} \, d\mu_i = \frac{2}{2 n+1} \biggl(\frac{\delta_{k,n-1} (1-n) (n+m)!}{(2 n-1) (n-m-1)!} + \frac{\delta_{k,n+1} (n+2) (n+m+1)!}{(2 n+3) (n-m)!}  \biggr) , \label{IntLeg2}\\
& \int_{-1}^1 \biggl((1-\mu_i^2) \frac{d P_n^m(\mu_i)}{d \mu_i} \frac{d P_k^m(\mu_i)}{d \mu_i} + \frac{m^2 P_n^m(\mu_i)P_k^m(\mu_i)}{1-\mu_i^2}\biggr) \mu_i \, d\mu_i \notag \\
&\qquad = \frac{2}{2 n+1}\biggl(\frac{\delta_{k,n-1} (n-1) (n+1) (n+m)!}{(2 n-1) (n-m-1)!} + \frac{\delta_{k,n+1} n (n+2) (n+m+1)!}{(2 n+3) (n-m)!}  \biggr) , \label{IntLeg3} \\
& \int_{-1}^1 P_n^m(\mu_i) P_k^{m-1}(\mu_i) (1-\mu_i^2)^{1/2} d\mu_i = \frac{2}{2 n+1} \biggl(\frac{-\delta_{k,n-1} (n+m)!}{(2 n-1) (n-m)!} + \frac{\delta_{k,n+1} (n+m)!}{(2 n+3) (n-m)!}  \biggr), \label{IntLeg4} \\
& \int_{-1}^1 P_n^m(\mu_i) P_k^{m+1}(\mu_i) (1-\mu_i^2)^{1/2} d\mu_i = \frac{2}{2 n+1} \biggl(\frac{\delta_{k,n-1} (n+m)!}{(2 n-1) (n-m-2)!} + \frac{-\delta_{k,n+1} (n+m+2)!}{(2 n+3) (n-m)!}  \biggr), \label{IntLeg5} \\
& \int_{-1}^1 \biggl((1-\mu_i^2)^{3/2} \frac{d P_n^m(\mu_i)}{d \mu_i} \frac{d P_k^{m-1}(\mu_i)}{d \mu_i} + \frac{m (m-1) P_n^m(\mu_i)P_k^{m-1}(\mu_i)}{(1-\mu_i^2)^{1/2}}\biggr) d\mu_i \notag \\
& \qquad = \frac{2}{2 n+1}\biggl(\frac{-\delta_{k,n-1} (n-1) (n+1) (n+m)!}{(2 n-1) (n-m)!} + \frac{\delta_{k,n+1} n (n+2) (n+m)!}{(2 n+3) (n-m)!}  \biggr) , \label{IntLeg6} \\
& \int_{-1}^1 \biggl((1-\mu_i^2)^{3/2} \frac{d P_n^m(\mu_i)}{d \mu_i} \frac{d P_k^{m+1}(\mu_i)}{d \mu_i} + \frac{m (m+1) P_n^m(\mu_i)P_k^{m+1}(\mu_i)}{(1-\mu_i^2)^{1/2}}\biggr) d\mu_i \notag \\
& \qquad = \frac{2}{2 n+1}\biggl(\frac{\delta_{k,n-1} (n-1) (n+1) (n+m)!}{(2 n-1) (n-m-2)!} + \frac{-\delta_{k,n+1} n (n+2) (n+m+2)!}{(2 n+3) (n-m)!}  \biggr) , \label{IntLeg7} \\
& \int_{-1}^1 \biggl( \mu_i (1-\mu_i^2)^{1/2} P_n^m(\mu_i) \frac{d P_k^{m-1}(\mu_i)}{d \mu_i} + \frac{m-1}{(1-\mu_i^2)^{1/2}}P_n^m(\mu_i)P_k^{m-1}(\mu_i) \biggr) d\mu_i \notag \\ 
&\qquad = \frac{2}{2 n+1}  \biggl(\frac{\delta_{k,n-1} (1-n) (n+m)!}{(2 n-1) (n-m)!} + \frac{-\delta_{k,n+1} (n+2) (n+m)!}{(2 n+3) (n-m)!}  \biggr) , \label{IntLeg8}\\
& \int_{-1}^1 \biggl( \mu_i (1-\mu_i^2)^{1/2} P_n^m(\mu_i) \frac{d P_k^{m+1}(\mu_i)}{d \mu_i} - \frac{m+1}{(1-\mu_i^2)^{1/2}}P_n^m(\mu_i)P_k^{m+1}(\mu_i) \biggr) d\mu_i \notag \\
& \qquad = \frac{2}{2 n+1} \biggl(\frac{\delta_{k,n-1} (n-1) (n+m)!}{(2 n-1) (n-m-2)!} + \frac{\delta_{k,n+1} (n+2) (n+m+2)!}{(2 n+3) (n-m)!}  \biggr) . \label{IntLeg9}
\end{align}
\end{subequations}
\end{widetext}
Integral \eqref{IntLeg1} is well-known (see e.g.~\cite{Samaddar1,GradRyzh,Varsh}). Integral \eqref{IntLeg2} (which appears in \cite[Eq.~(40)]{Samaddar1} as well, but is given there incorrectly) can be easily obtained by virtue of relation \cite[Eq.~(8.731)]{GradRyzh} $\frac{d P_k^m(\mu_i)}{d\mu_i} = \frac{(k+1) P_k^m(\mu_i)\mu_i - (k+1-m) P_{k+1}^m(\mu_i)}{1-\mu_i^2}$ and the previous integral \eqref{IntLeg1}. Integral \eqref{IntLeg3} boils down to those of \eqref{IntLeg1} and \eqref{IntLeg2} after integration by parts its first addend and taking into account that $P_k^m(\mu_i)$ satisfies the Legendre differential equation (see \cite[Eq.~(3.9)]{Jack}) $\frac{d}{d\mu_i}\left((1-\mu_i^2)\frac{d}{d\mu_i}P_k^m(\mu_i)\right) = \left(\frac{m^2}{1-\mu_i^2} - k(k+1)\right)P_k^m(\mu_i)$. All other integrals \eqref{IntsLeg} can also be calculated using similar tricks and employing recurrence properties (see \cite{GradRyzh}) for associated Legendre polynomials; let us finally note that in most cases we were unable to calculate these integrals in a general form by computer algebra systems for abstract (not predetermined by numbers) index values.

\bibliography{LinPaper}

\end{document}